\documentclass[sigconf]{acmart}
\usepackage{microtype}
\usepackage{graphicx}
\usepackage{subfigure}
\usepackage{balance}
\usepackage{booktabs} 
\usepackage{amsthm}
\usepackage{array}
\usepackage{graphicx}
\usepackage{clrscode}
\usepackage{subfigure}
\usepackage{multirow}
\usepackage{multicol}
\usepackage{float}
\usepackage{color}
\usepackage{diagbox}
\usepackage{xcolor}
\usepackage{amsopn}
\usepackage{mathrsfs}
\usepackage{mathtools}
\usepackage{amsmath}
\usepackage{booktabs}
\usepackage{arydshln}
\usepackage{hyperref}
\usepackage{blkarray}
\usepackage{enumerate}
\usepackage{courier}
\usepackage{mathrsfs}
\usepackage{rotating}
\usepackage{bm}
\usepackage{subfigure}
\usepackage{array}
\usepackage{ragged2e}
\usepackage{hyperref}
\usepackage{amsmath}
\usepackage{threeparttable}

\usepackage[ruled,linesnumbered]{algorithm2e}

\SetKwComment{comment}{ $triangleright$ \ }{}
\theoremstyle{definition}

\theoremstyle{theorem}

\theoremstyle{proof}

\theoremstyle{remark}

\AtBeginDocument{%
  \providecommand\BibTeX{{%
    \normalfont B\kern-0.5em{\scshape i\kern-0.25em b}\kern-0.8em\TeX}}}
\setcopyright{acmcopyright}
\copyrightyear{2018}
\acmYear{2018}
\acmDOI{10.1145/1122445.1122456}

\acmConference[Woodstock '18]{Woodstock '18: ACM Symposium on Neural
  Gaze Detection}{June 03--05, 2018}{Woodstock, NY}
\acmBooktitle{Woodstock '18: ACM Symposium on Neural Gaze Detection,
  June 03--05, 2018, Woodstock, NY}
\acmPrice{15.00}
\acmISBN{978-1-4503-XXXX-X/18/06}

\begin{document}

\title{Sequential Recommendation with User Evolving \\Preference Decomposition}

\author{Weiqi Shao$^{2,3}$, Xu Chen$^{2,3}$, Long Xia$^{1}$, Jiashu Zhao$^{1}$, Dawei Yin$^{1}$}

\affiliation{
\institution{$^1$Baidu Inc., China}
\city{}
\country{}
}
\affiliation{
\institution{$^2$Beijing Key Laboratory of Big Data Management and Analysis Methods, Beijing, China}
\city{}
\country{}
} 
\affiliation{
\institution{$^3$Gaoling School of Artificial Intelligence, Renmin University of China, Beijing, China}
\city{}
\country{}
}

\begin{abstract}
Modeling user sequential behaviors has recently attracted increasing attention in the recommendation domain.
Existing methods mostly assume coherent preference in the same sequence.
However, user personalities are volatile and easily changed, and there can be multiple mixed preferences underlying user behaviors.
To solve this problem, in this paper, we propose a novel sequential recommender model via decomposing and modeling user independent preferences.
To achieve this goal, we highlight three practical challenges considering the inconsistent, evolving and uneven nature of the user behavior, which are seldom noticed by the previous work.
For overcoming these challenges in a unified framework, we introduce a reinforcement learning module to simulate the evolution of user preference.
More specifically, the action aims to allocate each item into a sub-sequence or create a new one according to how the previous items are decomposed as well as the time interval between successive behaviors.
The reward is associated with the final loss of the learning objective, aiming to generate sub-sequences which can better fit the training data.
We conduct extensive experiments based on {six} real-world datasets across different domains.
Comparing with the state-of-the-art methods, empirical studies manifest that our model can on average improve the performance by about {8.21\%, 10.08\%, 10.32\% and 9.82\% on the metrics of Precision, Recall, NDCG and MRR}, respectively.
\end{abstract}
\maketitle

\begin{figure}[t]
\centering
\setlength{\fboxrule}{0.pt}
\setlength{\fboxsep}{0.pt}
\fbox{
\includegraphics[width=.99\linewidth]{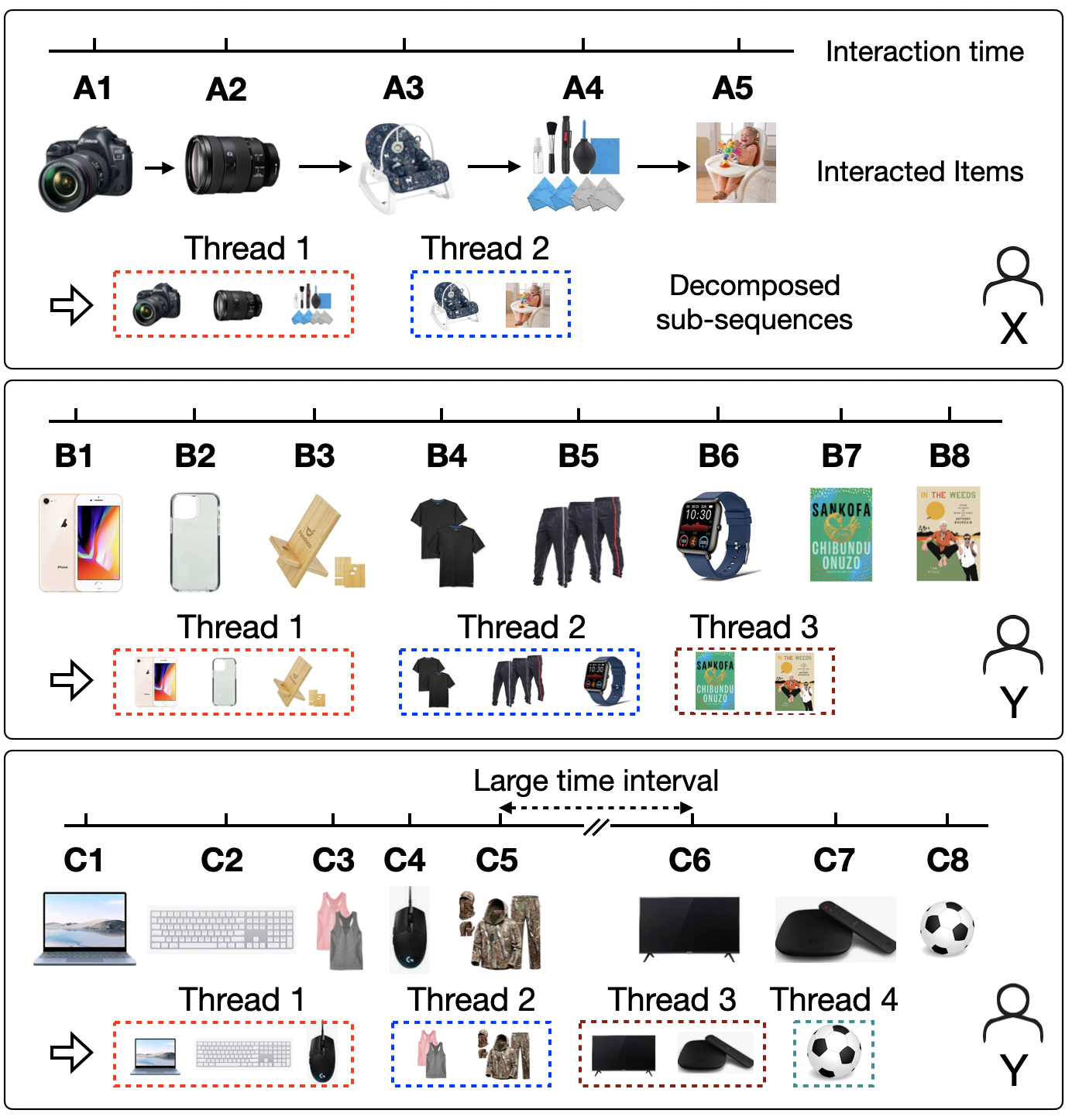}
}
\vspace*{-0.6cm}
\caption{
Examples of user inconsistent, evolving and uneven multi-thread preferences.
}
\label{intro-cx}
\vspace*{-0.6cm}
\end{figure}

\section{Introduction}\label{introduction}
Recommender system has been deployed in a wide range of applications, ranging from the fields of e-commerce~\cite{li2017neural,liu2018stamp,wang2019modeling}, education~\cite{lin2018intelligent,saito2020learning} to health-caring~\cite{zhou2020cnn,gong2021smr,bhoi2020premier} and entertainment~\cite{ayata2018emotion,subramaniyaswamy2019ontology,reddy2019content}.
Early recommender models like matrix factorization~\cite{he2017neural,xue2017deep,chen2020efficient} usually assume independence between different user behaviors.
However, user preference in real-world scenarios is an evolving process, and successive behaviors can be highly correlated.
For example, the purchasing of a mobile phone may trigger the interaction with the phone case, which may further attract user interest on phone holders.
Motivated by such characters, people have designed a lot of sequential recommender models~\cite{fang2020deep}. 
For example, FPMC~\cite{rendle2010factorizing} regards user behaviors as a Markov chain, where the current behavior is only influenced by the most recent action.
GRU4Rec~\cite{hidasi2015session} leverages recurrent neural network to summarize all the history behaviors for the next item prediction.

While sequential recommendation has achieved many promising results, existing methods usually assume coherent user preference in a sequence.
However, user personalities are complicated and diverse in practice, thus the same sequence may contain multiple user preferences. 
As exampled in the top block of Figure~\ref{intro-cx}, the user sequentially interacts with the items of ``camera $\rightarrow$ camera lens $\rightarrow$ baby chair $\rightarrow$ lens 
cleaner $\rightarrow$ hanging toy''. 
Their are two types of user preference.
One is on digital items, and the other is about baby products.
The user starts with the first preference, and purchases the ``camera'' and ``camera lens''.
And then, by interacting with the ``baby chair'', the user moves to the second preference.
In the next, the user returns to the first preference, and continues to buy the ``lens cleaner''.
At last, the second preference is triggered again via the interaction with the ``hanging toy''.
In this process, the user preference evolves along different threads.
For the digital items, the evolving path is ``camera $\rightarrow$ camera lens $\rightarrow$ lens cleaner''.
For the baby products, the user preference evolves from the ``baby chair'' to the ``hanging toy''.
Obviously, different preference threads have diverse evolving patterns, decomposing them and modeling each thread separately can lead to more clear history representation and better recommendation performance.
However, separating real-world user behaviors can be much more challenge because:

\textbf{CH1:} \textit{User behaviors are inconsistent.}
For different users, the number of preference threads may vary.
For example, in the top and middle blocks of Figure~\ref{intro-cx}, user X has two types of preferences on the digital and baby products, while the behaviors of user Y follow three preference threads along the book, sports and digital items, respectively.
Even for the same user, different behavior sequences may also have various number of preference threads, which is exampled in the middle and bottom blocks of Figure~\ref{intro-cx}.
In practice, how to handle such inconsistency across different sequences is challenging, since one cannot pre-define a unified thread number, and manually check each sequence is too labor intensive.

\textbf{CH2:} \textit{User behaviors are evolving processes.}
Straightforwardly, sequence decomposition can be seen as a classification problem, where one can predict the thread label for each item separately, and assemble the items with the same label into the final thread.
However, decomposing user behaviors is more complex.
See the example in the middle block of Figure~\ref{intro-cx}, in the beginning, the user interacted with many digital products.
And then, she moved to the sports items, and purchased the sport shirt, sweatpants and smart-watch.
If we classify the smart-watch independently, it may belong to the digital products, while by taking the user behaviors as an evolving process, we know that the reason of purchasing smart-watch is more likely for sports, such as timing and counting calories.
Such evolving nature is important and unique for user behavior decomposition, but how to model it is still under-explored.

\textbf{CH3:} \textit{User behaviors are unevenly distributed on the timeline.}
Different from sequence decomposition tasks in other domains like NLP~\cite{brill1995transformation} and CV~\cite{del2003decomposition}, a unique character of our task is the significance of the time interval information between successive behaviors.
Intuitively, if two behaviors happen with a large time interval, then they may have less correlations, and should be decomposed into different threads.
In the bottom block of Figure~\ref{intro-cx}, while ``C1 $\rightarrow$ C2 $\rightarrow$ C4'' and ``C6$\rightarrow$ C7'' both reflect user preference on the digital items, they may happen independently, since their interaction times have a large gap.
How to model such temporal information is important yet not well studied.

In order to overcome the above problems, in this paper, we design a reinforcement learning (RL) method to decompose user evolving preferences.
The agent simulates the generation of user behaviors, which outputs a set of sub-sequences representing user multi-thread preferences.
At each step, it decides which sub-sequence the current item should be allocated to or creates a new sub-sequence.
The reward is associated with the loss of the learning objective, aiming to obtain the allocation schemes which can better fit the training data.
In addition, we introduce auxiliary rewards to encourage that the items in a sub-sequence are coherent, while the representations of different sub-sequences are as independent as possible. 

We have noticed that there are some studies on multi-interest recommendation.
For example, SUM~\cite{lian2021multi} and LimaRec~\cite{wu2021rethinking} leverage attention model to determine item allocations.
MIND~\cite{li2019multi} and ComiRec~\cite{cen2020controllable} use capsule network and dynamic routing to separate different sub-sequences.
However, most of these models are not fully aware of the above challenges (\emph{i.e.}, CH1 to CH3), which may lead to sub-optimal model designs and lowered recommendation performance.
For clearly understanding the differences between our models and these previous work, we compare them in Table~\ref{intro-compare}.

In a summary, the main contributions of this paper can be concluded as follows:

$\bullet$ We propose to build sequential recommender models via decomposing user independent preferences, where we highlight three practical challenges brought by the inconsistent, evolving and uneven nature of the user behaviors.

$\bullet$ We design a reinforcement learning model to seamlessly overcome the above challenges in a unified framework, where we adaptively separate the complete user behavior sequence into many independent sub-sequences for better fitting the training data.

$\bullet$ We conduct extensive experiments based on {six} real-world datasets to demonstrate the superiority of our model.

\begin{table}[t]
\centering
\caption{{Comparisons between multi-interest and our models on solving the challenges mentioned in the introduction.}}
\vspace*{-0.2cm}
\scalebox{1.}{
\begin{tabular}
{
p{2.0cm}<{\centering}|
p{1.6cm}<{\centering}|
p{1.6cm}<{\centering}|
p{1.6cm}<{\centering}}
\hline\hline
{Model}  &CH1 &CH2& CH3\\ \hline
SUM~\cite{lian2021multi}&-&-&-\\ 
SINE~\cite{tan2021sparse}&-&-&-\\
DMIN~\cite{xiao2020deep}&-&-&-\\
MDSR~\cite{chen2021multi}&-&-&-\\
ComiRec~\cite{cen2020controllable}&-&-&-\\
MCPRN~\cite{wang2019modeling}&-&-&-\\
MIND~\cite{li2019multi}&\checkmark&-&-\\
Octopus~\cite{liu2020octopus}&\checkmark&-&-\\
PIMI~\cite{chen2021exploring}&-&-&\checkmark\\\hline
Our Model&\checkmark&\checkmark&\checkmark\\ \hline\hline
\end{tabular}
}
\label{intro-compare}
\vspace*{-0.5cm}
\end{table}

\section{{Sequential Recommendation Recapitulation}}
In sequential recommendation, the current user behavior is predicted by taking its history information into consideration.
Formally, we have a user set $\mathcal{U}$ and an item set $\mathcal{V}$.
The interactions of each user $u\in \mathcal{U}$ are chronologically organized into a set $\mathcal{O}_u = \{(v^u_1,t^u_1),(v^u_2,t^u_2),...,(v^u_{l_u},t^u_{l_u})\}$, where $t^u_i$ is the interaction time of item $v^u_i\in \mathcal{V}$, and $l_u$ is the number of interacted items. 
We denote by $\mathcal{O} = \{\mathcal{O}_u\}$ the set of all user-item interactions.
Then given $\{\mathcal{U},\mathcal{V},\mathcal{O}\}$, sequential recommendation aims to learn a model $f$, which can accurately predict the next item $v^u_{l_u+1}$ for user $u$ based on $\mathcal{O}_u$ and $t^u_{l_u+1}$.
It should be noted that, as a common simplification, the time information can be omitted~\cite{li2017neural,sun2019bert4rec,kang2018self}.

In the training process, $\mathcal{O}_u$ is recurrently separated into many samples $\{H^u_{j+1}, v^u_{j+1}\}_{j=1}^{l_u-1}$, where $H^u_{j+1} \!=\! [u, (v^u_1,t^u_1),...,(v^u_{j},t^u_{j}),t^u_{j+1}]$. 
Then, the following binary cross-entropy loss is used to learn $f$:
{\setlength\abovedisplayskip{5pt}
\setlength\belowdisplayskip{5pt}
\begin{eqnarray}\label{sin-seq}
\begin{aligned}
L_{1} =\!-\!\!\sum_{u=1}^{|\mathcal{U}|} \sum_{j=1}^{l_u-1}  [\log f(H^u_{j+1}, v^u_{j+1}) +\!\!\!\!\!\!\!\! \sum_{k\in S^-(v^u_{j+1})}\!\!\!\!\!\! \log(1-f(H^u_{j+1}, v^u_{k}))] 
\end{aligned}
\end{eqnarray}
}
where the output of $f(H^u_{j+1}, v^u_{j+1})$ is the probability of interacting with $v^u_{j+1}$ given the history information $H^u_{j+1}$.
$S^-(v^u_{j+1})$ is the set of negative samples, which can be sampled from the non-interacted items.
This objective aims to maximize the interaction probability of the real user behaviors, and simultaneously minimize that of the negative samples.

In the past few years, people have designed a lot of sequential recommender models~\cite{wang2021survey} to implement $f$.
However, most of them assume user preference to be coherent in a sequence, which is less reasonable given the potentially diverse user personalities.
To address this above problem, recent years have witnessed many multi-interest sequential recommender models (MIRM)~\cite{wang2019modeling,xiao2020deep,chen2021exploring,lian2021multi,wu2021rethinking}.
However, as mentioned before, these models fully or partially ignore the challenges of CH1 to CH3 (see Table~\ref{intro-compare}), which are key to decompose user behaviors in real-world scenarios.
By designing models tailored for these challenges, we propose a \underline{\textbf{s}}equential recommender model with ada\underline{\textbf{p}}tive user evo\underline{\textbf{l}}v\underline{\textbf{i}}ng preference decomposi\underline{\textbf{t}}ion (called SPLIT for short), which significantly differs from the previous work.

\begin{figure}[t]
\centering
\setlength{\fboxrule}{0.pt}
\setlength{\fboxsep}{0.pt}
\fbox{
\includegraphics[width=1.\linewidth]{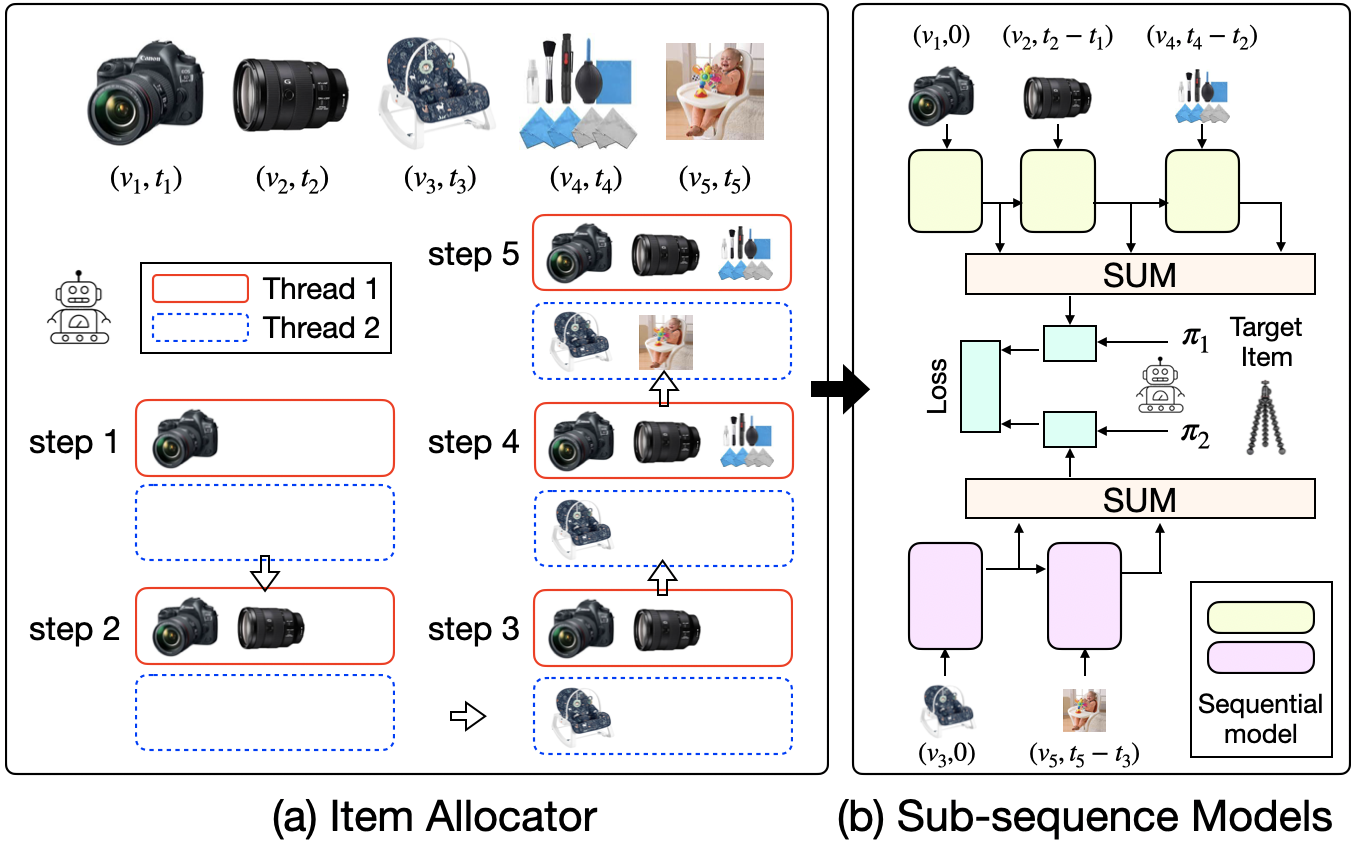}
}
\vspace*{-0.5cm}
\caption{
Illustration of our model: 
(a) is the item allocator, which sequentially assigns the items into different sub-sequences.
(b) is the sub-sequence modelers, which process the decomposed user behaviors from (a) to generate the final recommendation list.
}
\label{model}
\vspace*{-0.4cm}
\end{figure}

\section{THE SPLIT MODEL}
There are two major components in our model (see Figure~\ref{model}).
The first one is an item allocator, which aims to project the items into different preference threads.
The second one is a set of sub-sequence\footnote{From now on, we interchangeably use the terms of ``sub-sequence'' and ``preference thread''.} modelers, which are leveraged to handle the decomposed user behaviors.
In the serving process, the item allocator firstly decompose the complete sequence, and then the final recommendation is produced by pooling the results from the sub-sequence modelers.
In the following, we introduce these components more in detail.

\subsection{Item Allocator}\label{s3.1}
In most multi-interest sequential recommender models~\cite{li2019multi,cen2020controllable,chen2021multi}, the sub-sequences are generated by processing each item independently.
However, as analyzed above, the observed user behaviors are the evolution results of user multi-thread preferences. 
In order to accurately simulate the evolution process, we regard the decomposition of user preference as a Markov decision process (MDP), and design an RL based item allocator to separate user behavior sequences.
Generally speaking, we let the agent go through the target sequence, and at each step, the agent allocates the current item into an existing sub-sequence or create a new one. 
Our final goal is to obtain a set of sub-sequences, which can better fit the training data.
Formally, suppose we have a training sample $(H_{T+1}, v_{T+1})$, where $H_{T+1} = [u,(v_1,t_1),(v_2,t_2),...,(v_{T},t_{T}),t_{T+1}]$. Then, our task is to decompose $H_{T+1}$ into different sub-sequences\footnote{Here we omit the upper script $u$ when there is no confusion}, which can better predict $v_{T+1}$.
To achieve this goal, we define the following Markov decision process:

$\textbf{State}$ ($\bm{st}_i$):
The state at step $i$ is $(\bm{s}_i, t_i)$. 
$\bm{s}_i$ is set of existing sub-sequences before time step $i$, that is,
{\setlength\abovedisplayskip{5pt}
\setlength\belowdisplayskip{5pt}
\begin{eqnarray}\label{state}
\begin{aligned}
\bm{s}_i \!=\! \{\bm{s}_{i,b}\}_{b=1}^{k} \!=\! \{[u,(v_{i,b,1},t_{i,b,1}),...,(v_{i,b,l_b},t_{i,b,l_b})]\}_{b=1}^{k}\nonumber
\end{aligned}
\end{eqnarray}
}
where 
$v_{i,b,1}$ is an item in $\{v_k\}_{k=1}^T$, and $t_{i,b,1}$ is the corresponding interaction time.
$k$ is the number of sub-sequences, which is initialized as 1, and gradually increased in the item allocation process.
$l_b$ is the number of items in sub-sequence $b$, satisfying $\sum_{b=1}^k l_b = i-1$.

$\textbf{Action}$ (${a}_i$):
The action at step $i$ determines how to allocate the current item $v_i$.
If $v_i$ is coherent with an existing sub-sequence, then it will be put into it.
Otherwise, the agent will create a new sub-sequence initialized by $v_i$.
Formally, the action space is $[1,2,...,k,k+1]$, where $a_i=b~(b\in[1,k])$ means allocating $v_i$ to sub-sequence $b$, and $a_i=k+1$ indicates creating a new sub-sequence.
Remarkably, such action design is the key to accurately simulate user evolving preference, where the user can not only continue her old preferences, but also can launch a new interest.
After taking action $a_i$, $\bm{s}_i$ is updated to $\bm{s}_{i+1}$ in a deterministic manner, that is:
{\setlength\abovedisplayskip{5pt}
\setlength\belowdisplayskip{5pt}
\begin{eqnarray}\label{state1}
\begin{aligned}
\bm{s}_{i+1} = 
    \begin{cases}
        \{\bm{s}_{i,1}, ... , \bm{s}_{i,b}\cup[(v_{i},t_{i})],...,\bm{s}_{i,k}\} \ ,& \text{if} \ a_i=b\in[1,k] \\
        \{\bm{s}_{i,1}, ... ,\bm{s}_{i,k}, [(v_{i},t_{i})]\} \ ,& \text{if} \ a_i=k+1 \\
    \end{cases}
\end{aligned}
\end{eqnarray}
}
The state at step $i+1$ is $\bm{st}_{i+1} = (\bm{s}_{i+1}, t_{i+1})$.

$\textbf{Agent}$ ($\pi$):
At step $i$, the agent outputs the action distribution for allocating $v_i$ given the state $\bm{st}_i$.
In specific, each sub-sequence in $\bm{s}_i$ is firstly processed by a sequential model $g_s$ to derive the sub-sequence representation, that is:
{\setlength\abovedisplayskip{5pt}
\setlength\belowdisplayskip{5pt}
\begin{eqnarray}\label{cb}
\begin{aligned}
\bm{c}_{i,b} = g_s(\bm{s}_{i,b})\nonumber
\end{aligned}
\end{eqnarray}
}
where we delay the specification of $g_s$ in the following sections.
Given $\bm{c}_{i,b}$, the coherent score between $v_i$ and the sub-sequence $\bm{s}_{i,b}$ is computed as:
{\setlength\abovedisplayskip{5pt}
\setlength\belowdisplayskip{5pt}
\begin{eqnarray}\label{rb}
\begin{aligned}
cs_b = \text{dist}([\bm{c}_{i,b}; \bm{e}_i])\cdot \tau(t_i-t_b),
\end{aligned}
\end{eqnarray}
}
where $\bm{e}_i$ is the embedding of $v_i$, $[\cdot;\cdot]$ is the concatenate operation.
``$\text{dist}$'' is implemented with a three-layer fully connected neural network, measuring the similarity between $v_i$ and $\bm{s}_{i,b}$ in semantic.
$\tau$ is a monotonically decreasing function.
$t_b$ represents the sub-sequence time, which is set as the interaction time of the last item in $\bm{s}_{i,b}$ (\emph{i.e.}, $t_{i,b,l_b}$).
In this equation, we derive the overall coherent score based on two aspects, that is, the semantic similarity and the temporal influence.
Since $\tau$ is monotonically decreasing, the larger the time interval $(t_i-t_b)$ is, the smaller the overall coherent score $cs_b$ is.
This design encodes the intuition that the behaviors with larger time intervals have less correlations.

In ordinary reinforcement learning method, the action space is fixed.
However, an important aspect for simulating user evolving preference is modeling the emergence of new preference threads. 
To satisfy this requirement, we design a simple but effective method to adaptively expand the action space.
In specific, we introduce a threshold $\epsilon$, and define a novel activation function ``$\epsilon-$softmax'' to implement $\pi$, that is:
{\setlength\abovedisplayskip{5pt}
\setlength\belowdisplayskip{5pt}
\begin{eqnarray}\label{pi}
\begin{aligned}
\pi(a_i = a|(\bm{s}_i, t_i)) &= [\mu({cs}_1,{cs}_2,...,{cs}_k,\epsilon;\zeta)]_{a}\nonumber
\end{aligned}
\end{eqnarray}
}
where $\mu$ is an ordinary softmax operator with $\zeta$ as the temperature parameter. 
$[\bm{x}]_{a}$ selects the $a$th element of $\bm{x}$.
$a\in [1,k+1]$.
The working principle of $\epsilon-$softmax is as follows:
if $v_i$ is not coherent with any of existing sub-sequences, that is, ${cs}_b < \epsilon, \forall b\in[1,k]$, then the user is more likely to launch a new preference thread, correspondingly, $\arg\max_{a\in [1,k+1]} \pi(a_i = a|(\bm{s}_i, t_i)) = k+1$.
If there are many sub-sequences, satisfying ${cs}_b > \epsilon$, then $v_i$ will be allocated to the one with the largest ${cs}_b$, which agrees with our expectation, i.e., allocating $v_i$ to the most coherent sub-sequence.
As a special case, if ${cs}_b > \epsilon, \forall b\in[1,k]$, then our $\epsilon-$softmax is reduced to the ordinary softmax.
With $\epsilon-$softmax, the number of sub-sequences can be adaptively increased, which is able to capture the evolving nature of user preferences.

$\textbf{Reward}$ (${r}_i$):
We design rewards based on the following aspects:

$\bullet$ Whether the decomposed sub-sequences can lead to the better fitting of the training data?
In general, a model with lower loss means it can fit the data better.
Thus, we use the negative loss as the reward to measure the capability of data fitting.
For the sample $(H_{T+1}, v_{T+1})$, we denote by $l(H_{T+1}, v_{T+1})$ the loss function of predicting $v_{T+1}$ based on $H_{T+1}$.
Since we can only compute the loss when all the sub-sequences have been generated, the reward is defined in a delayed manner, that is:
{\setlength\abovedisplayskip{5pt}
\setlength\belowdisplayskip{5pt}
\begin{eqnarray}\label{r1}
\begin{aligned}
r_{i,1} = 
    \begin{cases}
        0                                    \ ,& \text{if} \ i\in[1,T-1] \\
        - l(H_{T+1}, v_{T+1})  \ ,& \text{if} \ i=T 
    \end{cases}
\end{aligned}
\end{eqnarray}
}

$\bullet$ Whether the allocated items can lead to better coherence within each sub-sequence?
To compute the sub-sequence coherence, we firstly compute the embedding of each sub-sequence $\bm{s}_{i,j}$ as the sum of its item embeddings, that is, $\bm{\kappa}_j = \frac{1}{|\bm{s}_{i,j}|}\sum_{k:(k,v)\in \bm{s}_{i,j}}\bm{e}_k$.
Suppose the action at step $i$ is $a$, $\bm{e}_i$ is the embedding of item $v_i$, then the reward is set as the cosine similarity between $\bm{s}_{i,b}$ and $v_i$, that is, 
{\setlength\abovedisplayskip{5pt}
\setlength\belowdisplayskip{5pt}
\begin{eqnarray}\label{r2}
\begin{aligned}
r_{i,2} = \frac{\bm{\kappa}_a^T\cdot \bm{e}_i}{||\bm{\kappa}_a||_2 ||\bm{e}_i||_2},
\end{aligned}
\end{eqnarray}
}
where $||\cdot||_2$ is the $L_2$-norm. If $a=k+1$, we directly set the reward as the similarity between the user and item embeddings.

$\bullet$ Whether the allocated items can lead to better orthogonality between different sub-sequences?
Ideally, each sub-sequence should exactly encode one type of user preference, thus we encourage orthogonality between different sub-sequences for avoiding information leaking.
Formally, if the action at step $i$ is $a$, then the reward is set as:
{\setlength\abovedisplayskip{5pt}
\setlength\belowdisplayskip{5pt}
\begin{eqnarray}\label{r3}
\begin{aligned}
r_{i,3} = -\sum_{a\ne j} \frac{||\hat{\bm{\kappa}}_a^T\cdot \bm{\kappa}_j||_2}{||\hat{\bm{\kappa}}_a||_2 ||\bm{\kappa}_j||_2},
\end{aligned}
\end{eqnarray}
}
where $\hat{\bm{\kappa}}_a = \frac{1}{|\bm{s}_{i,a}|+1}(\bm{e}_i + |\bm{s}_{i,a}| \bm{\kappa}_a)$ is the embedding of $\bm{s}_{i,a}$ after incorporating $v_i$.
Since only sub-sequence $a$ has been changed at this step, we only check the orthogonality between $\bm{s}_{i,a}$ and the other sub-sequences.

$\bullet$ Whether the number of sub-sequences is appropriate?
Intuitively, if the number of sub-sequences is too small, then user preferences may not be well separated.
However, if there are too many sub-sequences, then each sub-sequence may only contain a few items, which can be hard to represent the user preference.
In order to better control the sub-sequence number, we introduce the following reward: 
{\setlength\abovedisplayskip{5pt}
\setlength\belowdisplayskip{5pt}
\begin{equation}\label{r4}
    r_{i,4} = 
    \begin{cases}
        0        \ , & \text{if} \ a\in[1,k] \\
        \lambda  \ , & \text{if} \ a=k+1
    \end{cases}
\end{equation}
}
where if the current action $a$ is ``creating a new sub-sequence'', then there will be a non-zero reward $\lambda$.
If $\lambda>0$, then the agent is encouraged to produce more sub-sequences, otherwise, ``creating new sub-sequences'' is penalized.

For better guiding the sub-sequence generation, we implement $\lambda$ in a curriculum learning manner.
More specifically, in the beginning of the sequence decomposition, there are only a few sub-sequences.
At this time, we do not impose much constraint on ``creating new sub-sequences'', and set $\lambda$ as a large value.
As more sub-sequences are generated, we would like to control the increasing speed of the sub-sequence number, and thus lower the value of $\lambda$.
To realize this idea, we set $\lambda = w_1i+w_2$,
where $i$ is the index of the action step.
$w_1$ and $w_2$ are hyper-parameters, and we set $w_1<0$ to ensure that $\lambda$ is monotonically decreasing w.r.t. the index $i$, that is, as the agent takes more actions, we impose more strict control on the number of sub-sequences.

By combining $r_{i,1}$ to $r_{i,4}$, the overall reward at step $i$ is:
{\setlength\abovedisplayskip{5pt}
\setlength\belowdisplayskip{5pt}
\begin{eqnarray}\label{over-r}
\begin{aligned}
r_{i} = \sum_{j=1}^4 r_{i,j}.
\end{aligned}
\end{eqnarray}
}
A complete running example of the above MDP can be seen in Figure~\ref{model}(a).
Following the common practice in RL, the agent $\pi$ is learned based on the following objective: $E_{\pi}[\sum_{i=1}^T \gamma^{i-1}r_{i}]$, where $\gamma$ is a predefined discount factor.
Usually, the expectation is computed by its Monte Carlo approximation, and we leverage ``log-trick''~\cite{silver2014deterministic} to derive the following tractable objective: 
{\setlength\abovedisplayskip{5pt}
\setlength\belowdisplayskip{5pt}
\begin{eqnarray}\label{pi-loss}
\begin{aligned}
L_{IA}(\Theta_{\pi}, \Theta_{e}) = \sum_{u=1}^{|\mathcal{U}|} \sum_{j=1}^{l_u-1} [R^u_{j}\sum_{i=1}^{j}\log \pi(a^u_i|(\bm{s}^u_i, t^u_i))]
\end{aligned}
\end{eqnarray}
}
where $R^u_j = \sum_{i=1}^{j} \gamma^{i-1}r^u_{i}$, and we recover label $u$ to indicate the user of the training sample.
$\Theta_{e}$ is the user/item embedding matrices.
$\Theta_{\pi}=\{\theta_{s},\theta_{dist}\}$ is the parameter set of the agent.
$\theta_{s}$ and $\theta_{dist}$ are the parameters related with $g_s$ and $\text{dist}(\cdot,\cdot)$, respectively.

\noindent
\textbf{Model Specification.}
In the above sections, we have detailed our general idea on building RL-based item allocator.
For clear presentation, previously, we have skipped the implementations of $g_s$ and $\tau$. 
Here in the following, we elaborate their realizations.

(\romannumeral1) For $g_s$, it aims to project each sub-sequence into an embedding.
Actually, there are a lot of potential architectures to realize this function.
In this paper, we leverage LSTM~\cite{graves2013generating}, GRU~\cite{cho2014properties} and Transformer~\cite{vaswani2017attention} to implement $g_s$.
Give a sub-sequence $\bm{s}_{i,b}=[u,(v_{i,b,1},t_{i,b,1}),...,(v_{i,b,l_b},t_{i,b,l_b})]$, the input of $g_s$ at each step is $(v_{i,b,j},t_{i,b,j}-t_{i,b,j-1})$, where if $j=1$, then the input is $(v_{i,b,1},0)$.
To derive the embedding of the input, we firstly project the time interval $(t_{i,b,j}-t_{i,b,j-1})$ with a linear operator, and then cancat it with the embedding of $v_{i,b,j}$. 
We initialize the first step of $g_s$ with the user embedding, and the model architectures follow the original papers.
At last, we compute the final sub-sequence embedding as the average of the hidden states.

(\romannumeral2) For $\tau$, it aims to monotonically project the time interval information into a decaying constant in equation~(\ref{rb}).
In our model, we use two methods to implement $\tau$, that is:
\begin{equation}
    \tau(t_i-t_b) = 
    \begin{cases}
        -\kappa_1(t_i-t_b) + \kappa_2  \ , & \text{linear method}\\
        e^{-\kappa_3(t_i-t_b)}        \ , & \text{exponential method} 
    \end{cases}
\end{equation}
where $\kappa_1>0$ and $\kappa_3>0$ are pre-defined hyper-parameters.

\subsection{Sub-sequence Modeler}
The second component of our model is a set of sub-sequence modelers (see Figure~\ref{model}(b)).
For each training sample $(H_{T+1}, v_{T+1})$, where $H_{T+1} = (\bm{s}_{T+1}, t_{T+1})$, we firstly decompose $\bm{s}_{T+1}$ into $K$ sub-sequences $\{\bm{s}_{T+1,b}\}_{b=1}^{K}$ based on the above item allocator.
Then a sequential model $g_x$ is leveraged to project each $\bm{s}_{T+1,b}~(b\in[1,K])$ into an embedding, where we implement $g_x$ with the same architecture as used for $g_s$ in section~\ref{s3.1}.
Suppose the output embedding of $g_x(\bm{s}_{T+1,b})$ is $\bm{x}_b$, then the learning objective is:
{\setlength\abovedisplayskip{5pt}
\setlength\belowdisplayskip{5pt}
\begin{eqnarray}\label{r3}
\begin{aligned}
l(H_{T+1}, v_{T+1}) = -\sum_{b=1}^{K+1}  \pi_b l_{ce}(\bm{x}_b, \bm{e}_{T+1})
\end{aligned}
\end{eqnarray}
}
where 
$\pi_b = \pi(a_{T+1} = b|(\bm{s}_{T+1}, t_{T+1}))$ is the probability of allocating $v_{T+1}$ into an existing sub-sequence (when $b\in [1,K]$) or creating a new one initialized by $v_{T+1}$ (when $b=K+1$).
$\bm{e}_{T+1}$ is the embedding of item $v_{T+1}$.
$l_{ce}$ is the binary cross entropy loss used in equation~(\ref{sin-seq}), that is, $l_{ce}(\bm{x}_b, \bm{e}_{T+1}) = \log \sigma(\bm{x}_b^T \bm{e}_{T+1})+ \sum_{k\in S^-(v_{T+1})} \log(1-\sigma(\bm{x}_b^T \bm{e}_{k}))$.\footnote{When $b=K+1$, we assign $\bm{x}_b$ with the user embedding.}
At last, the overall learning objective is derived by summing the losses of all the training samples, that is:
{\setlength\abovedisplayskip{5pt}
\setlength\belowdisplayskip{5pt}
\begin{eqnarray}\label{ss-loss}
\begin{aligned}
L_{SSM}(\Theta_{e}, \Theta_{x}) = \sum_{u=1}^{|\mathcal{U}|} \sum_{j=1}^{l_u-1} l(H^u_{j+1}, v^u_{j+1}) \nonumber
\end{aligned}
\end{eqnarray}
}
where $H^u_{j+1} = (\bm{s}^u_{j+1}, t^u_{j+1})$, and we recover label $u$ to indicate the training samples related with user $u$.
$\Theta_{x}$ is the set of parameters related with $g_x$.

\subsection{Learning Process}
The complete learning process of our framework is summarized in Algorithm~\ref{random-alg}.
For each training sample, the agent generates a trajectory as follows:
at each step, the agent firstly takes an action based on the state, and then the reward is obtained based on equation~(\ref{over-r}), where the loss function in $r_{i,1}$ is derived by fixing $g_x$ obtained from the last training batch.
At last, the state $\bm{st}_i$ is updated to $\bm{st}_{i+1}$ based on equation~(\ref{state1}).
After obtaining the trajectory, the agent is optimized based on objective~(\ref{pi-loss}).
Once the agent has generated trajectories for all the training samples in a batch, we update the parameters of $\Theta_s, \Theta_e, \Theta_x$ jointly based on the following objective:
{\setlength\abovedisplayskip{5pt}
\setlength\belowdisplayskip{5pt}
\begin{eqnarray}\label{final-loss}
\begin{aligned}
L=\alpha L_{IA}(\Theta_{\pi}, \Theta_e) + (1-\alpha) L_{SSM}(\Theta_{x}, \Theta_e),
\end{aligned}
\end{eqnarray}
}
where $\alpha$ is a parameter balancing different optimization targets.

\setlength{\textfloatsep}{0.01cm}
\begin{algorithm}[t] 
\caption{Learning algorithm of our model}
\label{random-alg} 
Indicate the number of training batches M.\\
Initialize the model parameters $\Theta_{\pi}, \Theta_e, \Theta_x$.\\
Initialize a tuning parameter $\alpha$.\\
\For{batch number in [0, M]}{
    \For{each sample $(H_{T+1}, v_{T+1})$ in the batch}{
        \For{i in [1, T]}{
            Select an action $a_i$ based on $\bm{st}_i=(\bm{s}_i, t_i)$ and $\pi$.\\
            Allocate $v_i$ into sub-sequence $a_i$.\\
            Obtain the reward $r_i$.\\
            Update the state to $\bm{st}_{i+1}=(\bm{s}_{i+1}, t_{i+1})$.\\
        }
        Update $\Theta_{\pi}, \Theta_e$ based on objective~(\ref{pi-loss}) and $\{(\bm{s}_i, t_i), a_i, r_i\}_{i=1}^T$. \\
    }
    Update $\Theta_{\pi}, \Theta_e, \Theta_x$ based on objective~(\ref{final-loss}).\\
}
\end{algorithm}

\begin{table}[t]
\centering
\small
\caption{Relation between our model and the previous work.}
\vspace*{-0.2cm}
\scalebox{1.}{
\begin{tabular}
{
p{4.2cm}<{\centering}|
p{1.5cm}<{\centering}|
p{1.5cm}<{\centering}}
\hline\hline
{Model}  &Global &Local\\ \hline
General recommendation&Static&-\\ 
Sequential recommendation&Evolving&-\\
Multi-interest recommendation&Evolving&Static\\ 
SPLIT (our model)&Evolving&Evolving\\\hline\hline
\end{tabular}
}
\vspace*{-0.cm}
\label{model-compare-cx}
\end{table}

\subsection{Further Discussion}

\textbf{How do we overcome the challenges of CH1 to CH3?}
Our model is an effective remedy for the three challenges proposed in the introduction.
For \textbf{CH1}, we develop a novel $\epsilon$-softmax operator, which allows the number of user preference threads to be adaptively determined according to the target sequence.
For \textbf{CH2}, reinforcement learning is naturally designed for modeling the evolution of sequential events. 
In our model, the agent $\pi$ simulates how the user preference evolves, and the current item is allocated into a preference thread based on the previously generated sub-sequences.
For \textbf{CH3}, we incorporate the time information into the agent by equation~(\ref{rb}), where if the current item is far from a sub-sequence, then it is less likely to be allocated into it.

In fact, the above challenges are hard to be simultaneously addressed by existing multi-interest recommender models, which are mostly based on the capsule network and dynamic routing. In this paper, we follow a fundamentally different principle, and solve the above challenges seamlessly in a unified RL framework.

\noindent
\textbf{How does our model relate with the previous work?}
In general recommendation, the user-item interactions are modeled independently in a global manner.
We call this method as a ``globally static'' paradigm.
In sequential recommendation, the user evolving preference is considered, and all the items are modeled by a unified sequential model, which can be seen as a ``globally evolving'' paradigm.
In multi-interest recommendation, user evolving preference is modeled in a finer-grained manner, but the items are independently allocated into the sub-sequences.
We name this method as ``globally evolving and locally static'' paradigm.
In our model, we further improve multi-interest recommendation by capturing user evolving preference when generating sub-sequences, which is a ``globally evolving and locally evolving'' paradigm.
We summarize the above comparisons in Table~\ref{model-compare-cx} to clearly position our work, where we can see our model bridges an evident gap and has its unique contributions.

\section{Experiments}

\subsection{Experiment Setup}

\begin{table}[t]
\centering
\caption{{Statistics of the datasets used in our experiments.}}
\vspace{-0.2cm}
\scalebox{.83}{
\begin{tabular}
{
p{2.2cm}<{\centering}|
p{.8cm}<{\centering}|
p{.8cm}<{\centering}|
p{1.5cm}<{\centering}|
p{.9cm}<{\centering}|
p{1.7cm}<{\centering}}\hline\hline

{Dataset}     &{\#User}   &{\#Item}&{\#Interaction}  &{Density}&{Domain}   \\ \hline
{{Amazon-Video}}&5,131&1,686&37,126&0.43\%&e-commence\\ 
{{Amazon-Garden}}&1,687&963&13,272&0.82\%&e-commence\\ 
{{Amazon-Music}}&1,430&901&10,261&0.80\%&e-commence\\ 
{{Wechat}}&4.364&10,654&198,702&0.43\%&micro-video\\ 
{{Foursquare-NY}}&1,063&3,896&40,825&0.99\%&check-in\\ 
{{Foursquare-TKY}}&2,288&7,056&128,530&0.80\%&check-in\\ \hline\hline
\end{tabular}
}
\label{rec-dataset}
\vspace{-0.cm}
\end{table}

\subsubsection{Datasets}
Our experiments are conducted based on {six} real-world datasets across three different domains.
More specifically, 
\textbf{Amazon-Video}, \textbf{Amazon-Garden} and \textbf{Amazon-Music}\footnote{http://jmcauley.ucsd.edu/data/amazon/} are e-commerce datasets collected from Amazon.com, containing user purchasing records in different product categories.
\textbf{Wechat}\footnote{https://algo.weixin.qq.com/} is a video dataset, which provides the watching behaviors of the users on the videos.
\textbf{Foursquare-NY}\footnote{https://sites.google.com/site/yangdingqi/home/foursquare-dataset} and \textbf{Foursquare-TKY}~\cite{yang2013sentiment,sklar2012recommending,ye2010location} are well-known recommendation datasets, containing user check-in information in New York and Tokyo spanning for about 10 months. 
The statistics of these datasets are summarized in Table~\ref{rec-dataset}, where we can see: our datasets can cover different characters, \emph{e.g.}, we not only experiment in the e-commerce domain, but also have video and check-in datasets, and the dataset density spans a wide range.
With these diverse datasets, we hope to demonstrate the generalization of our idea, and evaluate different models in a fair manner.

\subsubsection{Baselines}
In order to demonstrate the effectiveness of our model, we compare it with 11 representative baselines.
We introduce these baselines following a similar taxonomy as in Table~\ref{model-compare-cx}:

$\bullet$ For general recommendation, 
\textbf{BPR}~\cite{rendle2012bpr} is a well-known recommender model for capturing user implicit feedback.
\textbf{NCF}~\cite{he2017neural}  is a deep recommender model, which generalizes the traditional matrix factorization method for modeling user-item nonlinear relationship.

$\bullet$ For sequential recommendation,
\textbf{GRU4Rec}~\cite{hidasi2015session} is an early method for modeling user sequential behaviors based on GRU.
\textbf{STAMP}~\cite{liu2018stamp} and \textbf{NARM}~\cite{li2017neural} are attention based sequential recommender models, where the former puts more focus on the last interaction.
\textbf{BERT4Rec}~\cite{tang2018personalized} is a sequential recommender model based on the self-attention mechanism.
For fair comparison, we also have two baselines involving the user-item interaction time information.
In specific, \textbf{TLSTM}~\cite{zhu2017next} is a novel time-aware LSTM architecture, which is able to incorporate the time information for user behavior modeling.
\textbf{TASER}~\cite{ye2020time} is a self-attention based recommender model, where the item embeddings are enhanced by the continuous time information.

$\bullet$ For multi-interest recommendation,
\textbf{MCPRN}~\cite{wang2019modeling} is a well known multi-interest recommender model, which associates each item with a type of user preference based on the capsule network and dynamic routing.
\textbf{PIMI}~\cite{chen2021exploring} is a time-sensitive multi-interest recommender model, where the user preferences are decomposed by considering the time interval information between successive behaviors.
\textbf{Octopus}~\cite{liu2020octopus} is a recently proposed multi-interest recommender model, which designs a novel ``activation'' mechanism, and different sequence may activate various number of interests.

\begin{table*}[!t]
\caption{\small{Overall comparison between our model and the baselines. 
All the numbers are percentage values with "\%" omitted.
We remove the prefix of Foursquare-NY and Foursquare-TKY for saving the space.
The best performance is labeled by bold fonts. 
The improvement of our model against the best baseline is significant under paired-t test with ``$p<0.05$''.
}}
\center
\small
\renewcommand\arraystretch{1.05}
\vspace{-0.3cm}
\setlength{\tabcolsep}{5.1pt}
\begin{threeparttable}  
\scalebox{0.97}{

\begin{tabular}
{
p{1.9cm}<{\centering}|p{1.4cm}<{\centering}|
p{.8cm}<{\centering}p{.8cm}<{\centering}|
p{.9cm}<{\centering}
p{.8cm}<{\centering}p{.8cm}<{\centering}
p{.9cm}<{\centering}p{.8cm}<{\centering}
p{.8cm}<{\centering}|p{0.8cm}<{\centering}
p{0.8cm}<{\centering}
p{0.8cm}<{\centering}|p{0.8cm}<{\centering}
                      } \hline\hline 
\multicolumn{2}{c|}{Category} &  \multicolumn{2}{c|}{General Model} &  \multicolumn{6}{c|}{Sequential Model} & \multicolumn{3}{c|}{Multi-interest Model}& \multicolumn{1}{c}{Our}    \\ \hline 

\multicolumn{2}{c|}{Method}& BPR & NCF & GRU4Rec & STAMP&NARM&BERT4Rec&TLSTM&TASER&MCPRN&PIMI&Octopus&SPLIT \\ \hline\hline

{\multirow{4}{*}{Amazon-Garden}}
& \multicolumn{1}{c|}{Precision@5}               
& 0.56 & 0.55 & 1.36
&  1.20 & 1.59  & 0.94
& 1.45 & 1.20
 & 1.60 & 1.55 
&  1.30 & \textbf{1.71} 
 \\  &
\multicolumn{1}{c|}{Recall@5}               
& 2.79 & 2.73 & 6.82 
&  5.99 & 7.95   & 4.69 
& 7.24 & 5.97
 & 8.01 & 7.01 
&  6.52 & \textbf{8.54} 
  \\ 
   &
\multicolumn{1}{c|}{NDCG@5}               
& 1.72 & 1.69 & 4.62 
&  4.90 & 5.21   & 2.71 
& 5.24 & 4.12  & 5.46 & 4.36 
&  3.50& \textbf{5.92}  \\ 
& \multicolumn{1}{c|}{MRR@5}               
& 1.37 & 1.35 & 3.88
&  4.22 & 4.31  & 2.07 
& 4.59 & 2.84
& 4.62 & 3.94 
&  2.52 & \textbf{5.07} 
 \\ 
\hline
\hline

{\multirow{4}{*}{Amazon-Video}}
& \multicolumn{1}{c|}{Precision@5}               

& 2.58 & 2.52 & 2.97  & 2.97
&  3.06 & 1.92   & 3.11 
& 2.68 & 2.74 &  2.95 & 2.16 
 & \textbf{3.43}\\  &
\multicolumn{1}{c|}{Recall@5}               

& 12.88 & 8.18 & 14.83 
&  14.84 & 15.95  & 9.62 
& 15.54 & 13.39   & 13.72 & 9.81
&  11.59 & \textbf{17.13}\\ 

 &
\multicolumn{1}{c|}{NDCG@5}               
& 8.46 & 5.12 & 10.37
&  11.22 & 11.22  & 7.87 
& 11.44 & 8.01  & 10.07 & 9.65 
& 8.11 & \textbf{13.23} \\ 
& \multicolumn{1}{c|}{MRR@5}               

& 6.99 & 6.72 & 8.90 
&  9.88 & 9.88  & 5.85
& 7.74 & 6.25&  8.87 & 8.66
&  6.97  & \textbf{11.93} \\ 
\hline\hline

{\multirow{4}{*}{Amazon-Music}}
& \multicolumn{1}{c|}{Precision@5}               
& 1.05 & 1.06 & 0.91
&  1.02 & 0.91  & 0.77
& 1.08 & 0.88
 & 1.13 & 1.04
&  1.07 & \textbf{1.33} 
 \\  &
\multicolumn{1}{c|}{Recall@5}               
& 5.25 & 5.32 & 4.55 &5.11 & 4.55
&  3.85 & 5.39   & 4.41
& 5.67 & 5.18
 & 5.23 & \textbf{6.65} 
  \\ 
   &
\multicolumn{1}{c|}{NDCG@5}               
& 3.34 & 3.59 & 3.01
&  3.57 & 3.14  & 2.17
& 3.62 & 3.09
& 3.82 & 3.43
&  3.56 & \textbf{4.22}   \\ 
& \multicolumn{1}{c|}{MRR@5}               
& 2.72 & 3.03 & 2.51
&  3.06 & 2.68  & 1.63
& 3.04 & 2.66
& 3.21 & 2.87
&  2.91 & \textbf{3.42} 
 \\ 
\hline
\hline

{\multirow{4}{*}{Wechat}}
& \multicolumn{1}{c|}{Precision@5}               
& 0.26 & 0.24 & 0.50  & 0.47
& 0.57 & 0.66 & 0.63  & 0.66
& 0.61 & 0.59 & 0.36  & \textbf{0.73}
\\&
\multicolumn{1}{c|}{Recall@5}               
& 1.31 & 1.21 & 2.50  & 2.34
& 2.87  & 2.93  & 3.14
& 3.12 & 3.05 & 2.93  & 1.81 & \textbf{3.67}
\\
&
\multicolumn{1}{c|}{NDCG@5}               
& 0.77 & 0.78 & 1.48  & 1.47
& 1.78 & 1.85 & 1.91  & 1.81
& 1.92& 1.85 & 1.04  & \textbf{2.25}
\\  
& \multicolumn{1}{c|}{MRR@5}               
& 0.60 & 0.62 & 1.15  & 1.19
& 1.43 & 1.50 & 1.52  & 1.51
& 1.55 & 1.50 & 0.79  & \textbf{1.79}
\\  
\hline\hline

{\multirow{4}{*}{NY}} 
& \multicolumn{1}{c|}{Precision@5}               
& 1.32 & 1.28 & 1.26 
&  1.32 & 1.34   & 1.39 
& 1.39 & 1.43 
 &  1.39 & 1.37
&  1.28 & \textbf{1.45}\\  &
\multicolumn{1}{c|}{Recall@5}               
& 6.59 & 6.40 & 6.31 
&  6.59 & 6.69   & 6.97
& 6.97 & 7.02  
 & 6.97 & 6.87
&  6.40 & \textbf{7.25}
\\ 
&
\multicolumn{1}{c|}{NDCG@5}               
& 4.32 & 4.29 & 4.04 
&  4.11 & 4.55   & 4.58
& 4.58 & 4.69 &  4.69 & 4.54
&  4.29 & \textbf{4.93}
 \\ 
& \multicolumn{1}{c|}{MRR@5}               
& 3.59 & 3.59 & 3.29 
&  3.29 & 3.84  & 3.97
& 3.79 & 3.88 
& 3.97 & 3.77
&  3.59 & \textbf{4.16} \\ 
\hline\hline

{\multirow{4}{*}{TKY}} 
& \multicolumn{1}{c|}{Precision@5}               
& 1.44 & 1.62 & 2.16 
&  1.97 & 2.08  & 2.08 
& 1.92 & 1.76  &1.78 & 1.78 
&  1.53& \textbf{2.21} \\  &
\multicolumn{1}{c|}{Recall@5}               

& 7.36 & 8.33 & 9.78
&  9.84 & 10.41   & 10.41 
& 9.58 & 8.79  & 8.92 & 8.88
&  7.65  & \textbf{11.06} \\ 
&
\multicolumn{1}{c|}{NDCG@5}               

& 4.55 & 5.12 & 6.90 
&  6.77 & 7.06  & 7.00 
& 6.66 & 6.77   & 6.10 & 6.44 
&  5.38 & \textbf{7.42}\\ 
& \multicolumn{1}{c|}{MRR@5}               

& 3.88 & 4.15 & 6.12 
&  5.77 & 5.97   & 5.88 
& 5.70 & 6.01 &  5.18 & 5.64 
&  4.63 & \textbf{6.22} \\ 
\hline\hline

\end{tabular}%
}          
\end{threeparttable}    
\label{tab:ov-result}   
\vspace{-0.3cm}
\end{table*}

\subsubsection{Implementation details}
For each user behavior sequence, we use the {last and second last interactions} as the testing and validation sets, while the others are left for model training. 
We evaluate different models based on the metrics including Precision, Recall, NDCG and MRR, where the former two measure the overlap between the recommended items and the ground truth, while the latter two are ranking based metrics, and the higher ranked accurate predictions contribute more to the results.
In the experiments, five items are recommended from each model to compare with the ground truth.
The parameters in our model are determined based on grid search.
More specifically, the learning rate and user/item embedding size are tuned in the ranges of $[0.01,0.005,0.001]$ and $[64,128,256]$, respectively.
The batch size is determined in the range of $[256,512,1024]$.
The threshold $\epsilon$ is searched in $[0.0,0.1,0.2,0.3,0.4,0.5,0.6,0.7,0.8,0.9]$.
The curriculum parameters $w_1$ and $w_2$ are determined in the ranges of $[-0.1,-0.2,-0.3,-0.4,-0.5]$ and $[0.1,0.2,0.3,0.4,0.5]$, respectively.
{The tuning parameter $\alpha$ is empirically set as 0.5.}
For the baselines, we set the parameters as their default values reported in the original papers or tune them in the same ranges as our model's.

\subsection{Overall Performance}\label{overall-p}
In this section, we compare our model with the baselines, and the results are presented in Table~\ref{tab:ov-result}, where we can see:
(\romannumeral1) in general, sequential recommender models can achieve better performance than the non-sequential ones.
These observations agree with the previous studies, and manifest that explicitly modeling the correlations between user behaviors is indeed useful to improve the recommendation performance.
(\romannumeral2) Among different sequential recommender models, the attention based ones usually perform better than the non-attention model (\emph{i.e.}, GRU4Rec).
We speculate that the attention mechanism can effectively discriminate the importances of different items.
The information which are useful for the target item prediction is strengthened, while the noisy items are lower weighted.
Thus, attention based models can better represent the history information, and bring superior performances.
Among all the baselines, the multi-interest recommender model MCPRN can usually achieve the best performance, which verifies the effectiveness of decomposing and modeling user independent preferences.
(\romannumeral3) Our model can achieve the best performance, which is consistent on all the evaluation metrics and datasets.
In specific, our model can on average improve the best baselines by about 8.21\%, 10.08\%, 10.32\% and 9.82\% on the metrics of Precision, Recall, NDCG and MRR, respectively.
Comparing with sequential recommender models, we are able to decompose the mixed user preferences, which facilitate more clear history representation and discriminative inference.
Comparing with multi-interest recommender models, our method can simultaneously consider the inconsistent, evolving and uneven characters of the user preference, which are important for decomposing user behaviors in practice.
We speculate that such designs may successfully reveal the basic rules underlying user behaviors, which introduce beneficial inductive bias, and lead to better recommendation performance.

\begin{figure}[t]
\centering
\setlength{\fboxrule}{0.pt}
\setlength{\fboxsep}{0.pt}
\fbox{
\includegraphics[width=.9\linewidth]{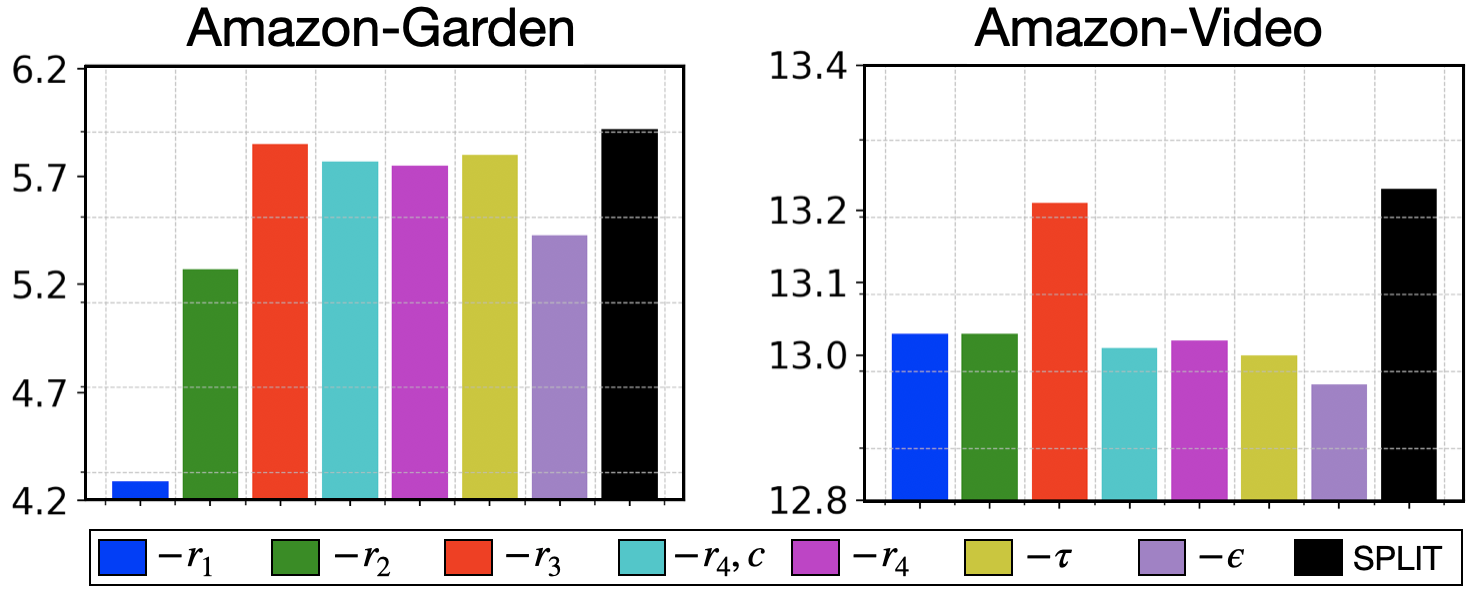}
}
\vspace*{-0.3cm}
\caption{
Results of the ablation studies. 
}
\label{ablation}
\vspace*{0.cm}
\end{figure}

\subsection{Ablation Studies}
In the above section, we compare our model with the baselines as a whole.
Readers may be interested in how different components in our model contribute the final performance.
To answer this question, we compare our model with its seven variants:
in \underline{\textbf{SPLIT (-$r_1$)}}, we remove the reward for better fitting the training data (\emph{i.e.}, equation~(\ref{r1})).
In \underline{\textbf{SPLIT (-$r_2$)}}, the reward for promoting sub-sequence coherence (\emph{i.e.}, equation~(\ref{r2})) is removed.
In \underline{\textbf{SPLIT (-$r_3$)}}, we do not impose constraints on the orthogonality between different sub-sequences, that is, we remove the reward of equation~(\ref{r3}).
In \underline{\textbf{SPLIT (-$r_4$, c)}}, we do not use curriculum learning in equation~(\ref{r4}), where we fix $\lambda$ as a hyper-parameter\footnote{In the experiments, we tune $\lambda$ to report the best performance.}.
In \underline{\textbf{SPLIT (-$r_4$)}}, we completely drop the reward of equation~(\ref{r4}).
In \underline{\textbf{SPLIT (-$\tau$)}}, we drop the temporal influence, where we remove $\tau(t_i-t_b)$ from equation~(\ref{rb}).
In \underline{\textbf{SPLIT (-$\epsilon$)}}, we implement our model with fixed number of user preference threads, where we set the thread number $K$ as a hyper-parameter and leverage ordinary softmax to implement $\pi$.
In the experiments, the model parameters follow the above settings, and we report the performance based on NDCG and the datasets of Amazon-Garden and Amazon-Video, respectively.
The results on the other metrics and datasets are similar and omitted.

From the results shown in Figure~\ref{ablation}, we can see:
the reward for better fitting the training data (\emph{i.e.}, $r_1$) is very important, which is evidenced by the lowered performance of SPLIT (-$r_1$) comparing with SPLIT.
On the dataset of Amazon-Garden, the performance of SPLIT (-$r_1$) is even worse than many baselines.
Actually, this is not surprising, since this reward is directly related with the loss function, which is critical for the model performance.
SPLIT (-$r_2$) and SPLIT (-$r_3$) perform worse than SPLIT on both datasets, which suggests that the reward on promoting intra-subsequence coherence (\emph{i.e.}, $r_2$) and inter-subsequence orthogonality (\emph{i.e.}, $r_3$) are both necessary.
It is very interesting to see that the performances of SPLIT (-$r_4$, c) and SPLIT (-$r_4$) are similar.
This manifests that if we control the number of sub-sequences in a too simple manner, \emph{e.g.}, defining an unchanged reward, then its effect is similar to that of not setting this reward at all.
By carefully designing the reward function, SPLIT is better than both SPLIT (-$r_4$, c) and SPLIT (-$r_4$), which demonstrates the effectiveness of our proposed curriculum reward.
By comparing SPLIT (-$\tau$) with SPLIT, we find that the temporal function $\tau$ is important, since removing it can lead to lowered performances on both datasets.
This result actually verifies the effectiveness of addressing the third challenge mentioned in the introduction, and confirms the importance of modeling the time interval information between successive user behaviors.
If we unify the number of sub-sequences for all the samples, the performance is not satisfied, especially, for Amazon-Video, SPLIT (-$\epsilon$) is the worst among all the variants.
This observation highlights the significance of overcoming the first challenge in the introduction, which suggests that modelling user behavior inconsistency is important.

\subsection{Further Experiments}
In this section, we conduct further experiments to study the influence of different model implementations and the key hyper-parameters.
Similar to the above settings, we base the experiments on the datasets of Amazon-Garden and Amazon-Video, and use NDCG as the evaluation metric.
The results on the other datasets and metrics are similar and omitted.
We set the model parameters as their optimal values tuned in section~\ref{overall-p}.

\begin{table}[t]
\centering
\caption{{Influence of different $g_s$'s and $\tau$'s.}}
\vspace{-0.3cm}
\scalebox{1.}{
\begin{tabular}
{
p{2.1cm}<{\centering}|p{1.4cm}<{\centering}|p{1.4cm}<{\centering}|p{1.7cm}<{\centering}}\hline\hline
\multicolumn{4}{c}{Amazon-Garden}   \\ \hline
\diagbox[width=2.3cm, height=.5cm, trim=l]{$\tau$}{$g_s$}&{LSTM}&{GRU}&{Transformer} \\ \hline
Linear &8.88&13.00&11.49\\ \hline
Exponential &6.89&13.32&13.13\\ \hline\hline
\multicolumn{4}{c}{Amazon-Video}   \\ \hline
\diagbox[width=2.3cm, height=.5cm, trim=l]{$\tau$}{$g_s$}&{LSTM}&{GRU}&{Transformer} \\ \hline
Linear &2.78&5.4&5.41\\ \hline
Exponential &3.24&5.92&5.46\\ \hline\hline
\end{tabular}
}
\vspace*{-0.cm}
\label{fa1}
\end{table}

\subsubsection{Influence of different $g_s$'s and $\tau$'s}
In our model, $g_s$ determines how to represent the generated sub-sequences\footnote{It should be noted that we implement $g_x$ and $g_s$ with the same architecture, thus we do not discuss different implementations of $g_x$.}, and $\tau$ indicates how the time interval information influence the coherence score.
In this paper, we explore three methods to implement $g_s$, that is: LSTM, GRU and transformer.
For $\tau$, we use either linear or exponential methods to project the time interval information into a constant.
In the experiment, the hyper-parameters $\kappa_1$, $\kappa_2$ and $\kappa_3$ are tuned to achieve the best performance.
We compare different combinations between $g_s$ and $\tau$ in Table~\ref{fa1}, where we can see:
for the sequential architecture $g_s$, GRU can usually lead to the best performance.
We speculate that GRU is a much lighter architecture comparing with LSTM and transformer, which can be more appropriate for the sparse recommendation datasets.
More redundant parameters may over-fit the training data and lead to inferior performances.
From the perspective of $\tau$, exponential function can be the better choice for our task in most cases.
The reason can be that the exponential method can project any time-interval information into a moderate range (\emph{i.e,}, [0,1]), while the linear method may result in too large $\tau(t-t_b)$'s, which may overwhelm the semantic similarity in equation~(\ref{rb}), and impact the final performance.

\subsubsection{Influence of the threshold $\epsilon$}
\begin{figure}[t]
\centering
\setlength{\fboxrule}{0.pt}
\setlength{\fboxsep}{0.pt}
\fbox{
\includegraphics[width=.95\linewidth]{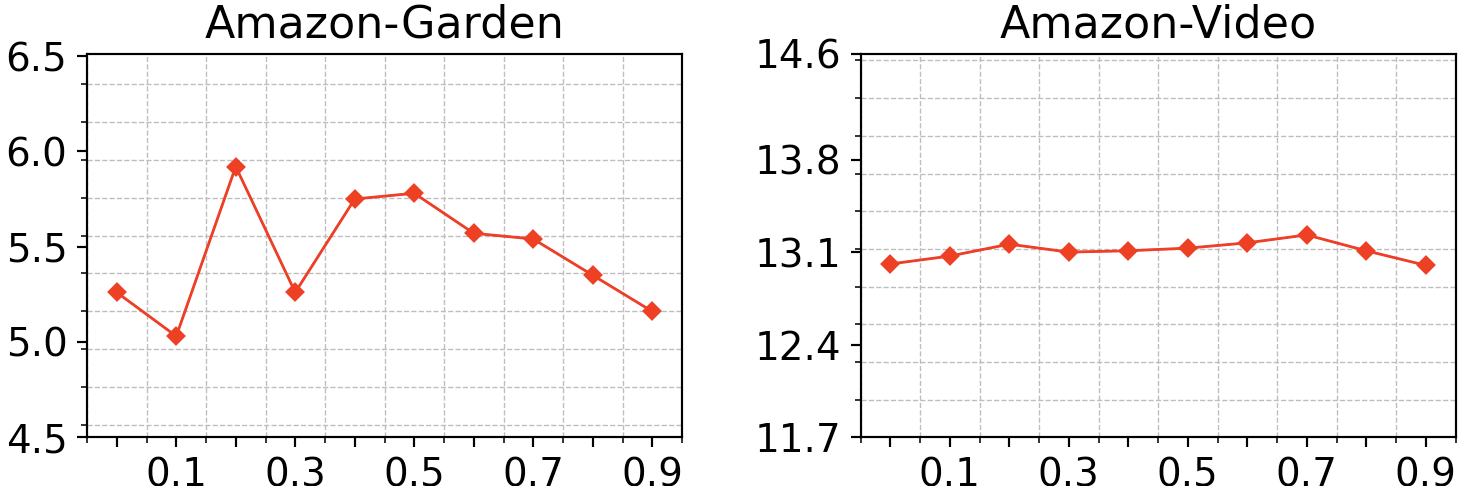}
}
\vspace*{-0.3cm}
\caption{
Influence of the threshold $\epsilon$.
}
\label{thres}
\vspace*{-0.cm}
\end{figure}

In our agent $\pi$, the threshold $\epsilon$ determines how rigorous we would like to constrain the generation of a new sub-sequence.
With a small $\epsilon$, the agent can not easily produce new sub-sequences, since the condition ``$cs_b < \epsilon$'' is hard to satisfy.
While if $\epsilon$ is larger, then the agent has more changes to generate new sub-sequences. 
In order to see the influence of $\epsilon$ on the final performance, we tune it in the range of $[0.0,0.1,0.2,0.3,0.4,0.5,0.6,0.7,0.8,0.9]$, and the results are presented in Figure~\ref{thres}, where we can see:
on Amazon-Garden, the performance fluctuates dramatically when $\epsilon$ is small, and after reaching the optimal point, the performance continually goes down as $\epsilon$ becomes larger.
On Amazon-Video, the performance curve is much smoother, but similarly, it has experienced the ``going-up'' and ``going-down'' processes.
On both datasets, the best performance is achieved when $\epsilon$ is moderate.
We speculate that too large $\epsilon$ makes the agent produce too many sub-sequences, and there is only a few items in each sub-sequence, which is limited for comprehensively representing the user preference.
While if $\epsilon$ is too small, the user diverse preferences cannot be well decomposed, and the mixed preferences may confuse the sub-sequence representations and lower the model performance.

\subsubsection{Influence of the curriculum parameters $w_1$ and $w_2$}
\begin{figure}[t]
\centering
\setlength{\fboxrule}{0.pt}
\setlength{\fboxsep}{0.pt}
\fbox{
\includegraphics[width=1.\linewidth]{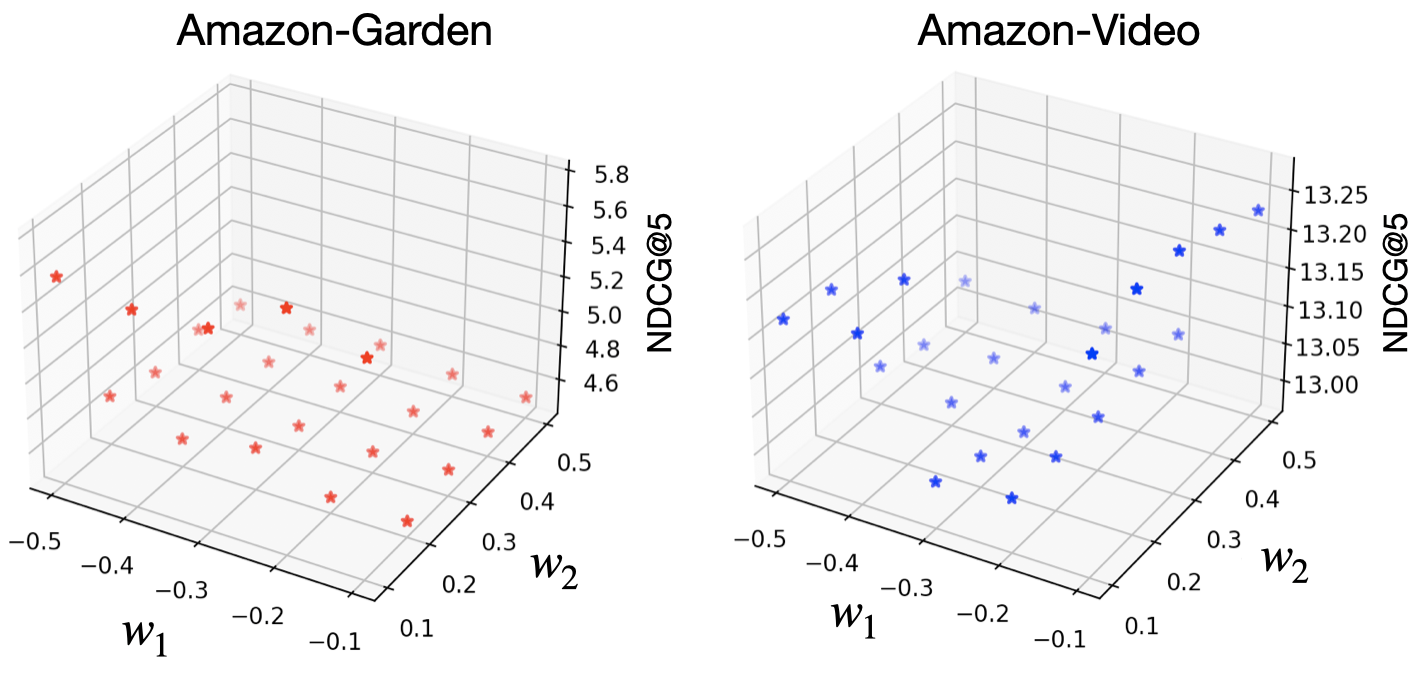}
}
\vspace*{-0.5cm}
\caption{
Influence of $w_1$ and $w_2$.
}
\label{3d}
\vspace*{-0.cm}
\end{figure}

In reward $r_{i,4}$ (\emph{i.e.}, equation~(\ref{r4})), the number of user preference threads is controlled in a curriculum learning manner.
$w_1$ and $w_2$ determine the curves of the reward.
In specific, $w_2$ sets the initial reward at the first step, while $w_1<0$ indicates the reward decreasing speed.
In this section, we study the influence of $w_1$ and $w_2$ on the final performance.
We tune $w_1$ and $w_2$ in the ranges of $[-0.1,-0.2,-0.3,-0.4,-0.5]$ and $[0.1,0.2,0.3,0.4,0.5]$, respectively, and the results are shown in Figure~\ref{3d}, from which we can see: on the dataset of Amazon-Garden, smaller $w_1$ and $w_2$ can usually lead to better performances, while on Amazon-Video, the best performance is achieved when $w_1$ and $w_2$ are larger.
We speculate that the number of items in Amazon-Garden is small, thus their characters can be easily covered by a few amount of independent user preference threads.
To limit the thread number, smaller $w_1$ and $w_2$ can help to set a lower initial reward on ``encouraging new sub-sequences'', and decrease this reward sharply in the following action steps.
However, in Amazon-Video, there are more items, which needs more user preference threads to cover their properties, thus $w_1$ and $w_2$ should be set as larger values to generate more sub-sequences.

\subsection{Case Studies}
\begin{figure}[t]
\centering
\setlength{\fboxrule}{0.pt}
\setlength{\fboxsep}{0.pt}
\fbox{
\includegraphics[width=.97\linewidth]{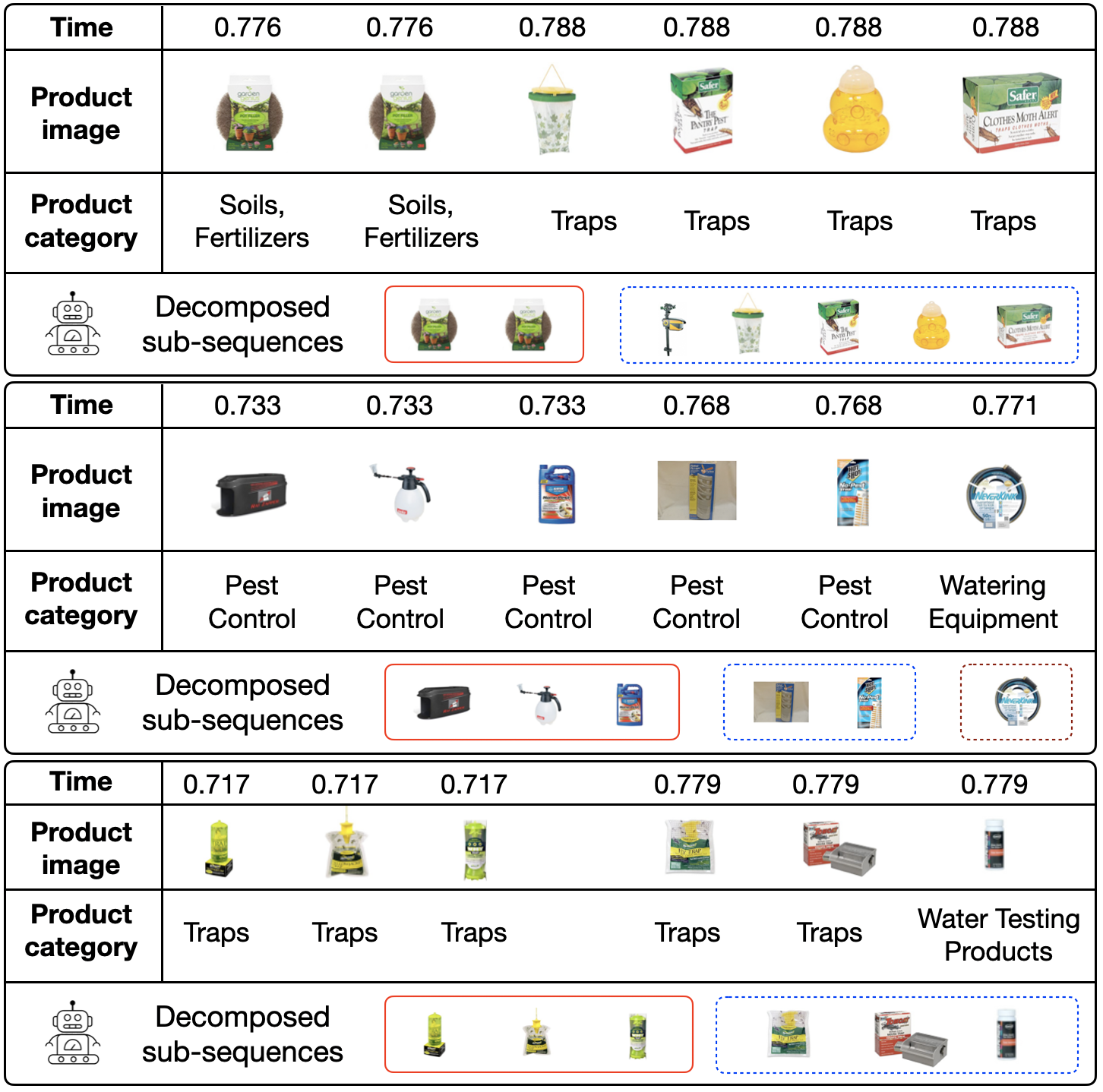}
}
\vspace*{-0.2cm}
\caption{
Case studies on the sub-sequences, where the time information is normalized into a value between 0 and 1.
}
\label{case}
\vspace*{-0.cm}
\end{figure}

In order to provide more intuitive understandings on the decomposed user behaviors, in this section, we analyze our model by presenting several case studies.
We experiment with the dataset of Amazon-Garden, and the model parameters follow their optimal values determined in section~\ref{overall-p}.
For each case, we decompose the complete user behavior sequence based on the learned item allocator.
From the results presented in Figure~\ref{case}, we can see:
(\romannumeral1) the products with different semantics can be successfully separated into different sub-sequences.
For example, in the first case, the items on fertilizers are classified into the first sub-sequence, while the trap products are decomposed into the second one.
(\romannumeral2) For different sequences, the item allocator can adaptively separate it into different numbers of sub-sequences.
For example, in the first and second cases, there are two and three sub-sequences, respectively.
(\romannumeral3) The temporal information can indeed influence the sequence decomposition.
For example, in the third case, while the first five products are all traps, the time interval between the third and forth ones are very large, thus they are separated into different sub-sequences. Similar result can also be found in the second case.

\section{Related work}
\subsection{Sequential Recommendation}
Our paper is targeted at the problem of sequential recommendation~\cite{wang2021survey}.
Early methods~\cite{rendle2010factorizing} in this field regard user behaviors as Markov chains, and the influence from the history behaviors is assumed to concentrate on the latest actions.
While these methods are efficient due to dropping the user behaviors longer before, their performances are sub-optimal.
A natural idea to solve this problem is increasing the history length.
Following this direction, people have proposed a number of models based on recurrent neural network~\cite{hidasi2015session,ma2019hierarchical}, convolutional neural network~\cite{tang2018personalized,yuan2019simple}, memory network~\cite{chen2018sequential} and transformer~\cite{sun2019bert4rec,kang2018self}.
A common character of these methods is that they can capture user long-term preference, and the prediction of the current item can be influenced by the behaviors happened longer before.
These models also have their respective advantages.
For example, many of them~\cite{sun2019bert4rec,kang2018self} can adaptively learn the importances of the previous items.
Some of them~\cite{chen2018sequential} can store information for extremely long sequence modeling.
The major difference between our model and these methods is that we decompose user mixed preferences in the behavior sequence, which facilitates more clear and focused prediction.

\subsection{Multi-interest Recommendation}
Our model is also related with multi-interest recommendation (MIR).
The general idea of MIR is to associate each item with a type of user interest (similar to the concept of ``thread'' in our formulation), and the final recommendation score is computed by pooling the similarities between the candidate item and the representation of each user interest.
The major difference of existing multi-interest recommender models lies in how to associate the items with the user interests, typical methods include computing user interests based attention scores~\cite{wang2019modeling,xiao2020deep,chen2021exploring,lian2021multi,wu2021rethinking} and leveraging dynamic routing to learn the associations between the items and user interests~\cite{li2019multi,cen2020controllable,chen2021multi}.
More specifically,
~\cite{xiao2020deep} extracts user independent preferences based on the self-attentive mechanism, where different attention heads are regarded as the representations of various user interests.
~\cite{wang2019modeling,cen2020controllable,chen2021multi} leverage capsule network and dynamic routing to associate each item with the candidate user preferences.
~\cite{chen2021exploring} incorporates temporal information to derive the item embeddings for multi-interest decomposition.
While these models have achieved many promising results, as mentioned before, most of them fully or partially ignore the three key challenges proposed in the introduction, which is important for decomposing real-world user behaviors.

\section{Conclusions}
In this paper, we highlight the significance of building sequential recommender models via decomposing user evolving preference.
We propose three key challenges one needs to face for achieving this goal, and design an RL based model to seamlessly overcome these challenges.
Extensive empirical studies manifest that our model can outperform a series of state-of-the-art methods, and such superiority is consistent across different recommendation domains.
We believe this work makes a novel step towards capturing user finer-grained evolving preference, and there is much room left for improvement.
To begin with, we may incorporate more comprehensive side information to provide more valuable signals for allocating the current item into a sub-sequence.
In addition, one may also extend our idea to the graph setting, where the user preferences can be decomposed into various sub-graphs.

\bibliographystyle{ACM-Reference-Format}
\balance
\bibliography{acmart}


\begin{thebibliography}{45}


\ifx \showCODEN    \undefined \def \showCODEN     #1{\unskip}     \fi
\ifx \showDOI      \undefined \def \showDOI       #1{#1}\fi
\ifx \showISBNx    \undefined \def \showISBNx     #1{\unskip}     \fi
\ifx \showISBNxiii \undefined \def \showISBNxiii  #1{\unskip}     \fi
\ifx \showISSN     \undefined \def \showISSN      #1{\unskip}     \fi
\ifx \showLCCN     \undefined \def \showLCCN      #1{\unskip}     \fi
\ifx \shownote     \undefined \def \shownote      #1{#1}          \fi
\ifx \showarticletitle \undefined \def \showarticletitle #1{#1}   \fi
\ifx \showURL      \undefined \def \showURL       {\relax}        \fi
\providecommand\bibfield[2]{#2}
\providecommand\bibinfo[2]{#2}
\providecommand\natexlab[1]{#1}
\providecommand\showeprint[2][]{arXiv:#2}

\bibitem[\protect\citeauthoryear{Ayata, Yaslan, and Kamasak}{Ayata
  et~al\mbox{.}}{2018}]%
        {ayata2018emotion}
\bibfield{author}{\bibinfo{person}{Deger Ayata}, \bibinfo{person}{Yusuf
  Yaslan}, {and} \bibinfo{person}{Mustafa~E Kamasak}.}
  \bibinfo{year}{2018}\natexlab{}.
\newblock \showarticletitle{Emotion based music recommendation system using
  wearable physiological sensors}.
\newblock \bibinfo{journal}{\emph{IEEE transactions on consumer electronics}}
  \bibinfo{volume}{64}, \bibinfo{number}{2} (\bibinfo{year}{2018}),
  \bibinfo{pages}{196--203}.
\newblock


\bibitem[\protect\citeauthoryear{Bhoi, Li, and Hsu}{Bhoi et~al\mbox{.}}{2020}]%
        {bhoi2020premier}
\bibfield{author}{\bibinfo{person}{Suman Bhoi}, \bibinfo{person}{Lee~Mong Li},
  {and} \bibinfo{person}{Wynne Hsu}.} \bibinfo{year}{2020}\natexlab{}.
\newblock \showarticletitle{Premier: Personalized recommendation for medical
  prescriptions from electronic records}.
\newblock \bibinfo{journal}{\emph{arXiv preprint arXiv:2008.13569}}
  (\bibinfo{year}{2020}).
\newblock


\bibitem[\protect\citeauthoryear{Brill}{Brill}{1995}]%
        {brill1995transformation}
\bibfield{author}{\bibinfo{person}{Eric Brill}.}
  \bibinfo{year}{1995}\natexlab{}.
\newblock \showarticletitle{Transformation-based error-driven learning and
  natural language processing: A case study in part-of-speech tagging}.
\newblock \bibinfo{journal}{\emph{Computational linguistics}}
  \bibinfo{volume}{21}, \bibinfo{number}{4} (\bibinfo{year}{1995}),
  \bibinfo{pages}{543--565}.
\newblock


\bibitem[\protect\citeauthoryear{Cen, Zhang, Zou, Zhou, Yang, and Tang}{Cen
  et~al\mbox{.}}{2020}]%
        {cen2020controllable}
\bibfield{author}{\bibinfo{person}{Yukuo Cen}, \bibinfo{person}{Jianwei Zhang},
  \bibinfo{person}{Xu Zou}, \bibinfo{person}{Chang Zhou},
  \bibinfo{person}{Hongxia Yang}, {and} \bibinfo{person}{Jie Tang}.}
  \bibinfo{year}{2020}\natexlab{}.
\newblock \showarticletitle{Controllable multi-interest framework for
  recommendation}. In \bibinfo{booktitle}{\emph{Proceedings of the 26th ACM
  SIGKDD International Conference on Knowledge Discovery \& Data Mining}}.
  \bibinfo{pages}{2942--2951}.
\newblock


\bibitem[\protect\citeauthoryear{Chen, Zhang, Zhang, Liu, and Ma}{Chen
  et~al\mbox{.}}{2020}]%
        {chen2020efficient}
\bibfield{author}{\bibinfo{person}{Chong Chen}, \bibinfo{person}{Min Zhang},
  \bibinfo{person}{Yongfeng Zhang}, \bibinfo{person}{Yiqun Liu}, {and}
  \bibinfo{person}{Shaoping Ma}.} \bibinfo{year}{2020}\natexlab{}.
\newblock \showarticletitle{Efficient neural matrix factorization without
  sampling for recommendation}.
\newblock \bibinfo{journal}{\emph{ACM Transactions on Information Systems
  (TOIS)}} \bibinfo{volume}{38}, \bibinfo{number}{2} (\bibinfo{year}{2020}),
  \bibinfo{pages}{1--28}.
\newblock


\bibitem[\protect\citeauthoryear{Chen, Zhang, Zhao, Xue, and Xiang}{Chen
  et~al\mbox{.}}{2021b}]%
        {chen2021exploring}
\bibfield{author}{\bibinfo{person}{Gaode Chen}, \bibinfo{person}{Xinghua
  Zhang}, \bibinfo{person}{Yanyan Zhao}, \bibinfo{person}{Cong Xue}, {and}
  \bibinfo{person}{Ji Xiang}.} \bibinfo{year}{2021}\natexlab{b}.
\newblock \showarticletitle{Exploring Periodicity and Interactivity in
  Multi-Interest Framework for Sequential Recommendation}.
\newblock \bibinfo{journal}{\emph{arXiv preprint arXiv:2106.04415}}
  (\bibinfo{year}{2021}).
\newblock


\bibitem[\protect\citeauthoryear{Chen, Ren, Cai, Sun, and De~Rijke}{Chen
  et~al\mbox{.}}{2021a}]%
        {chen2021multi}
\bibfield{author}{\bibinfo{person}{Wanyu Chen}, \bibinfo{person}{Pengjie Ren},
  \bibinfo{person}{Fei Cai}, \bibinfo{person}{Fei Sun}, {and}
  \bibinfo{person}{Maarten De~Rijke}.} \bibinfo{year}{2021}\natexlab{a}.
\newblock \showarticletitle{Multi-interest Diversification for End-to-end
  Sequential Recommendation}.
\newblock \bibinfo{journal}{\emph{ACM Transactions on Information Systems
  (TOIS)}} \bibinfo{volume}{40}, \bibinfo{number}{1} (\bibinfo{year}{2021}),
  \bibinfo{pages}{1--30}.
\newblock


\bibitem[\protect\citeauthoryear{Chen, Xu, Zhang, Tang, Cao, Qin, and Zha}{Chen
  et~al\mbox{.}}{2018}]%
        {chen2018sequential}
\bibfield{author}{\bibinfo{person}{Xu Chen}, \bibinfo{person}{Hongteng Xu},
  \bibinfo{person}{Yongfeng Zhang}, \bibinfo{person}{Jiaxi Tang},
  \bibinfo{person}{Yixin Cao}, \bibinfo{person}{Zheng Qin}, {and}
  \bibinfo{person}{Hongyuan Zha}.} \bibinfo{year}{2018}\natexlab{}.
\newblock \showarticletitle{Sequential recommendation with user memory
  networks}. In \bibinfo{booktitle}{\emph{Proceedings of the eleventh ACM
  international conference on web search and data mining}}.
  \bibinfo{pages}{108--116}.
\newblock


\bibitem[\protect\citeauthoryear{Cho, Van~Merri{\"e}nboer, Bahdanau, and
  Bengio}{Cho et~al\mbox{.}}{[n.\,d.]}]%
        {cho2014properties}
\bibfield{author}{\bibinfo{person}{Kyunghyun Cho}, \bibinfo{person}{Bart
  Van~Merri{\"e}nboer}, \bibinfo{person}{Dzmitry Bahdanau}, {and}
  \bibinfo{person}{Yoshua Bengio}.} \bibinfo{year}{[n.\,d.]}\natexlab{}.
\newblock \showarticletitle{On the properties of neural machine translation:
  Encoder-decoder approaches}.
\newblock  (\bibinfo{year}{[n.\,d.]}).
\newblock


\bibitem[\protect\citeauthoryear{Del~Vecchio, Murray, and Perona}{Del~Vecchio
  et~al\mbox{.}}{2003}]%
        {del2003decomposition}
\bibfield{author}{\bibinfo{person}{Domitilla Del~Vecchio},
  \bibinfo{person}{Richard~M Murray}, {and} \bibinfo{person}{Pietro Perona}.}
  \bibinfo{year}{2003}\natexlab{}.
\newblock \showarticletitle{Decomposition of human motion into dynamics-based
  primitives with application to drawing tasks}.
\newblock \bibinfo{journal}{\emph{Automatica}} \bibinfo{volume}{39},
  \bibinfo{number}{12} (\bibinfo{year}{2003}), \bibinfo{pages}{2085--2098}.
\newblock


\bibitem[\protect\citeauthoryear{Fang, Zhang, Shu, and Guo}{Fang
  et~al\mbox{.}}{2020}]%
        {fang2020deep}
\bibfield{author}{\bibinfo{person}{Hui Fang}, \bibinfo{person}{Danning Zhang},
  \bibinfo{person}{Yiheng Shu}, {and} \bibinfo{person}{Guibing Guo}.}
  \bibinfo{year}{2020}\natexlab{}.
\newblock \showarticletitle{Deep learning for sequential recommendation:
  Algorithms, influential factors, and evaluations}.
\newblock \bibinfo{journal}{\emph{ACM Transactions on Information Systems
  (TOIS)}} \bibinfo{volume}{39}, \bibinfo{number}{1} (\bibinfo{year}{2020}),
  \bibinfo{pages}{1--42}.
\newblock


\bibitem[\protect\citeauthoryear{Gong, Wang, Wang, Wang, and Liu}{Gong
  et~al\mbox{.}}{2021}]%
        {gong2021smr}
\bibfield{author}{\bibinfo{person}{Fan Gong}, \bibinfo{person}{Meng Wang},
  \bibinfo{person}{Haofen Wang}, \bibinfo{person}{Sen Wang}, {and}
  \bibinfo{person}{Mengyue Liu}.} \bibinfo{year}{2021}\natexlab{}.
\newblock \showarticletitle{Smr: Medical knowledge graph embedding for safe
  medicine recommendation}.
\newblock \bibinfo{journal}{\emph{Big Data Research}}  \bibinfo{volume}{23}
  (\bibinfo{year}{2021}), \bibinfo{pages}{100174}.
\newblock


\bibitem[\protect\citeauthoryear{Graves}{Graves}{[n.\,d.]}]%
        {graves2013generating}
\bibfield{author}{\bibinfo{person}{Alex Graves}.}
  \bibinfo{year}{[n.\,d.]}\natexlab{}.
\newblock \showarticletitle{Generating sequences with recurrent neural
  networks}.
\newblock  (\bibinfo{year}{[n.\,d.]}).
\newblock


\bibitem[\protect\citeauthoryear{He, Liao, Zhang, Nie, Hu, and Chua}{He
  et~al\mbox{.}}{2017}]%
        {he2017neural}
\bibfield{author}{\bibinfo{person}{Xiangnan He}, \bibinfo{person}{Lizi Liao},
  \bibinfo{person}{Hanwang Zhang}, \bibinfo{person}{Liqiang Nie},
  \bibinfo{person}{Xia Hu}, {and} \bibinfo{person}{Tat-Seng Chua}.}
  \bibinfo{year}{2017}\natexlab{}.
\newblock \showarticletitle{Neural collaborative filtering}. In
  \bibinfo{booktitle}{\emph{Proceedings of the 26th international conference on
  world wide web}}. \bibinfo{pages}{173--182}.
\newblock


\bibitem[\protect\citeauthoryear{Hidasi, Karatzoglou, Baltrunas, and
  Tikk}{Hidasi et~al\mbox{.}}{2015}]%
        {hidasi2015session}
\bibfield{author}{\bibinfo{person}{Bal{\'a}zs Hidasi},
  \bibinfo{person}{Alexandros Karatzoglou}, \bibinfo{person}{Linas Baltrunas},
  {and} \bibinfo{person}{Domonkos Tikk}.} \bibinfo{year}{2015}\natexlab{}.
\newblock \showarticletitle{Session-based recommendations with recurrent neural
  networks}.
\newblock \bibinfo{journal}{\emph{arXiv preprint arXiv:1511.06939}}
  (\bibinfo{year}{2015}).
\newblock


\bibitem[\protect\citeauthoryear{Kang and McAuley}{Kang and McAuley}{2018}]%
        {kang2018self}
\bibfield{author}{\bibinfo{person}{Wang-Cheng Kang} {and}
  \bibinfo{person}{Julian McAuley}.} \bibinfo{year}{2018}\natexlab{}.
\newblock \showarticletitle{Self-attentive sequential recommendation}. In
  \bibinfo{booktitle}{\emph{2018 IEEE International Conference on Data Mining
  (ICDM)}}. IEEE, \bibinfo{pages}{197--206}.
\newblock


\bibitem[\protect\citeauthoryear{Li, Liu, Wu, Xu, Zhao, Huang, Kang, Chen, Li,
  and Lee}{Li et~al\mbox{.}}{2019}]%
        {li2019multi}
\bibfield{author}{\bibinfo{person}{Chao Li}, \bibinfo{person}{Zhiyuan Liu},
  \bibinfo{person}{Mengmeng Wu}, \bibinfo{person}{Yuchi Xu},
  \bibinfo{person}{Huan Zhao}, \bibinfo{person}{Pipei Huang},
  \bibinfo{person}{Guoliang Kang}, \bibinfo{person}{Qiwei Chen},
  \bibinfo{person}{Wei Li}, {and} \bibinfo{person}{Dik~Lun Lee}.}
  \bibinfo{year}{2019}\natexlab{}.
\newblock \showarticletitle{Multi-interest network with dynamic routing for
  recommendation at Tmall}. In \bibinfo{booktitle}{\emph{Proceedings of the
  28th ACM International Conference on Information and Knowledge Management}}.
  \bibinfo{pages}{2615--2623}.
\newblock


\bibitem[\protect\citeauthoryear{Li, Ren, Chen, Ren, Lian, and Ma}{Li
  et~al\mbox{.}}{2017}]%
        {li2017neural}
\bibfield{author}{\bibinfo{person}{Jing Li}, \bibinfo{person}{Pengjie Ren},
  \bibinfo{person}{Zhumin Chen}, \bibinfo{person}{Zhaochun Ren},
  \bibinfo{person}{Tao Lian}, {and} \bibinfo{person}{Jun Ma}.}
  \bibinfo{year}{2017}\natexlab{}.
\newblock \showarticletitle{Neural attentive session-based recommendation}. In
  \bibinfo{booktitle}{\emph{Proceedings of the 2017 ACM on Conference on
  Information and Knowledge Management}}. \bibinfo{pages}{1419--1428}.
\newblock


\bibitem[\protect\citeauthoryear{Lian, Batal, Liu, Soni, Kang, Wang, and
  Xie}{Lian et~al\mbox{.}}{2021}]%
        {lian2021multi}
\bibfield{author}{\bibinfo{person}{Jianxun Lian}, \bibinfo{person}{Iyad Batal},
  \bibinfo{person}{Zheng Liu}, \bibinfo{person}{Akshay Soni},
  \bibinfo{person}{Eun~Yong Kang}, \bibinfo{person}{Yajun Wang}, {and}
  \bibinfo{person}{Xing Xie}.} \bibinfo{year}{2021}\natexlab{}.
\newblock \showarticletitle{Multi-Interest-Aware User Modeling for Large-Scale
  Sequential Recommendations}.
\newblock \bibinfo{journal}{\emph{arXiv preprint arXiv:2102.09211}}
  (\bibinfo{year}{2021}).
\newblock


\bibitem[\protect\citeauthoryear{Lin, Pu, Li, and Lian}{Lin
  et~al\mbox{.}}{2018}]%
        {lin2018intelligent}
\bibfield{author}{\bibinfo{person}{Jinjiao Lin}, \bibinfo{person}{Haitao Pu},
  \bibinfo{person}{Yibin Li}, {and} \bibinfo{person}{Jian Lian}.}
  \bibinfo{year}{2018}\natexlab{}.
\newblock \showarticletitle{Intelligent recommendation system for course
  selection in smart education}.
\newblock \bibinfo{journal}{\emph{Procedia Computer Science}}
  \bibinfo{volume}{129} (\bibinfo{year}{2018}), \bibinfo{pages}{449--453}.
\newblock


\bibitem[\protect\citeauthoryear{Liu, Zeng, Mokhosi, and Zhang}{Liu
  et~al\mbox{.}}{2018}]%
        {liu2018stamp}
\bibfield{author}{\bibinfo{person}{Qiao Liu}, \bibinfo{person}{Yifu Zeng},
  \bibinfo{person}{Refuoe Mokhosi}, {and} \bibinfo{person}{Haibin Zhang}.}
  \bibinfo{year}{2018}\natexlab{}.
\newblock \showarticletitle{STAMP: short-term attention/memory priority model
  for session-based recommendation}. In \bibinfo{booktitle}{\emph{Proceedings
  of the 24th ACM SIGKDD International Conference on Knowledge Discovery \&
  Data Mining}}. \bibinfo{pages}{1831--1839}.
\newblock


\bibitem[\protect\citeauthoryear{Liu, Lian, Yang, Lian, and Xie}{Liu
  et~al\mbox{.}}{[n.\,d.]}]%
        {liu2020octopus}
\bibfield{author}{\bibinfo{person}{Zheng Liu}, \bibinfo{person}{Jianxun Lian},
  \bibinfo{person}{Junhan Yang}, \bibinfo{person}{Defu Lian}, {and}
  \bibinfo{person}{Xing Xie}.} \bibinfo{year}{[n.\,d.]}\natexlab{}.
\newblock \showarticletitle{Octopus: Comprehensive and Elastic User
  Representation for the Generation of Recommendation Candidates}.
\newblock  (\bibinfo{year}{[n.\,d.]}).
\newblock


\bibitem[\protect\citeauthoryear{Ma, Kang, and Liu}{Ma et~al\mbox{.}}{2019}]%
        {ma2019hierarchical}
\bibfield{author}{\bibinfo{person}{Chen Ma}, \bibinfo{person}{Peng Kang}, {and}
  \bibinfo{person}{Xue Liu}.} \bibinfo{year}{2019}\natexlab{}.
\newblock \showarticletitle{Hierarchical gating networks for sequential
  recommendation}. In \bibinfo{booktitle}{\emph{Proceedings of the 25th ACM
  SIGKDD international conference on knowledge discovery \& data mining}}.
  \bibinfo{pages}{825--833}.
\newblock


\bibitem[\protect\citeauthoryear{Reddy, Nalluri, Kunisetti, Ashok, and
  Venkatesh}{Reddy et~al\mbox{.}}{2019}]%
        {reddy2019content}
\bibfield{author}{\bibinfo{person}{SRS Reddy}, \bibinfo{person}{Sravani
  Nalluri}, \bibinfo{person}{Subramanyam Kunisetti}, \bibinfo{person}{S Ashok},
  {and} \bibinfo{person}{B Venkatesh}.} \bibinfo{year}{2019}\natexlab{}.
\newblock \showarticletitle{Content-based movie recommendation system using
  genre correlation}.
\newblock In \bibinfo{booktitle}{\emph{Smart Intelligent Computing and
  Applications}}. \bibinfo{publisher}{Springer}, \bibinfo{pages}{391--397}.
\newblock


\bibitem[\protect\citeauthoryear{Rendle, Freudenthaler, Gantner, and
  Schmidt-Thieme}{Rendle et~al\mbox{.}}{2012}]%
        {rendle2012bpr}
\bibfield{author}{\bibinfo{person}{Steffen Rendle}, \bibinfo{person}{Christoph
  Freudenthaler}, \bibinfo{person}{Zeno Gantner}, {and} \bibinfo{person}{Lars
  Schmidt-Thieme}.} \bibinfo{year}{2012}\natexlab{}.
\newblock \showarticletitle{BPR: Bayesian personalized ranking from implicit
  feedback}.
\newblock \bibinfo{journal}{\emph{arXiv preprint arXiv:1205.2618}}
  (\bibinfo{year}{2012}).
\newblock


\bibitem[\protect\citeauthoryear{Rendle, Freudenthaler, and
  Schmidt-Thieme}{Rendle et~al\mbox{.}}{2010}]%
        {rendle2010factorizing}
\bibfield{author}{\bibinfo{person}{Steffen Rendle}, \bibinfo{person}{Christoph
  Freudenthaler}, {and} \bibinfo{person}{Lars Schmidt-Thieme}.}
  \bibinfo{year}{2010}\natexlab{}.
\newblock \showarticletitle{Factorizing personalized markov chains for
  next-basket recommendation}. In \bibinfo{booktitle}{\emph{Proceedings of the
  19th international conference on World wide web}}. \bibinfo{pages}{811--820}.
\newblock


\bibitem[\protect\citeauthoryear{Saito and Watanobe}{Saito and
  Watanobe}{2020}]%
        {saito2020learning}
\bibfield{author}{\bibinfo{person}{Tomohiro Saito} {and}
  \bibinfo{person}{Yutaka Watanobe}.} \bibinfo{year}{2020}\natexlab{}.
\newblock \showarticletitle{Learning path recommendation system for programming
  education based on neural networks}.
\newblock \bibinfo{journal}{\emph{International Journal of Distance Education
  Technologies (IJDET)}} \bibinfo{volume}{18}, \bibinfo{number}{1}
  (\bibinfo{year}{2020}), \bibinfo{pages}{36--64}.
\newblock


\bibitem[\protect\citeauthoryear{Silver, Lever, Heess, Degris, Wierstra, and
  Riedmiller}{Silver et~al\mbox{.}}{2014}]%
        {silver2014deterministic}
\bibfield{author}{\bibinfo{person}{David Silver}, \bibinfo{person}{Guy Lever},
  \bibinfo{person}{Nicolas Heess}, \bibinfo{person}{Thomas Degris},
  \bibinfo{person}{Daan Wierstra}, {and} \bibinfo{person}{Martin Riedmiller}.}
  \bibinfo{year}{2014}\natexlab{}.
\newblock \showarticletitle{Deterministic policy gradient algorithms}. In
  \bibinfo{booktitle}{\emph{International conference on machine learning}}.
  PMLR, \bibinfo{pages}{387--395}.
\newblock


\bibitem[\protect\citeauthoryear{Sklar, Shaw, and Hogue}{Sklar
  et~al\mbox{.}}{2012}]%
        {sklar2012recommending}
\bibfield{author}{\bibinfo{person}{Max Sklar}, \bibinfo{person}{Blake Shaw},
  {and} \bibinfo{person}{Andrew Hogue}.} \bibinfo{year}{2012}\natexlab{}.
\newblock \showarticletitle{Recommending interesting events in real-time with
  foursquare check-ins}. In \bibinfo{booktitle}{\emph{Proceedings of the sixth
  ACM conference on Recommender systems}}. \bibinfo{pages}{311--312}.
\newblock


\bibitem[\protect\citeauthoryear{Subramaniyaswamy, Manogaran, Logesh,
  Vijayakumar, Chilamkurti, Malathi, and Senthilselvan}{Subramaniyaswamy
  et~al\mbox{.}}{2019}]%
        {subramaniyaswamy2019ontology}
\bibfield{author}{\bibinfo{person}{V Subramaniyaswamy},
  \bibinfo{person}{Gunasekaran Manogaran}, \bibinfo{person}{R Logesh},
  \bibinfo{person}{V Vijayakumar}, \bibinfo{person}{Naveen Chilamkurti},
  \bibinfo{person}{D Malathi}, {and} \bibinfo{person}{N Senthilselvan}.}
  \bibinfo{year}{2019}\natexlab{}.
\newblock \showarticletitle{An ontology-driven personalized food recommendation
  in IoT-based healthcare system}.
\newblock \bibinfo{journal}{\emph{The Journal of Supercomputing}}
  \bibinfo{volume}{75}, \bibinfo{number}{6} (\bibinfo{year}{2019}),
  \bibinfo{pages}{3184--3216}.
\newblock


\bibitem[\protect\citeauthoryear{Sun, Liu, Wu, Pei, Lin, Ou, and Jiang}{Sun
  et~al\mbox{.}}{2019}]%
        {sun2019bert4rec}
\bibfield{author}{\bibinfo{person}{Fei Sun}, \bibinfo{person}{Jun Liu},
  \bibinfo{person}{Jian Wu}, \bibinfo{person}{Changhua Pei},
  \bibinfo{person}{Xiao Lin}, \bibinfo{person}{Wenwu Ou}, {and}
  \bibinfo{person}{Peng Jiang}.} \bibinfo{year}{2019}\natexlab{}.
\newblock \showarticletitle{BERT4Rec: Sequential recommendation with
  bidirectional encoder representations from transformer}. In
  \bibinfo{booktitle}{\emph{Proceedings of the 28th ACM international
  conference on information and knowledge management}}.
  \bibinfo{pages}{1441--1450}.
\newblock


\bibitem[\protect\citeauthoryear{Tan, Zhang, Yao, Liu, Zhou, Yang, and Hu}{Tan
  et~al\mbox{.}}{2021}]%
        {tan2021sparse}
\bibfield{author}{\bibinfo{person}{Qiaoyu Tan}, \bibinfo{person}{Jianwei
  Zhang}, \bibinfo{person}{Jiangchao Yao}, \bibinfo{person}{Ninghao Liu},
  \bibinfo{person}{Jingren Zhou}, \bibinfo{person}{Hongxia Yang}, {and}
  \bibinfo{person}{Xia Hu}.} \bibinfo{year}{2021}\natexlab{}.
\newblock \showarticletitle{Sparse-interest network for sequential
  recommendation}. In \bibinfo{booktitle}{\emph{Proceedings of the 14th ACM
  International Conference on Web Search and Data Mining}}.
  \bibinfo{pages}{598--606}.
\newblock


\bibitem[\protect\citeauthoryear{Tang and Wang}{Tang and Wang}{2018}]%
        {tang2018personalized}
\bibfield{author}{\bibinfo{person}{Jiaxi Tang} {and} \bibinfo{person}{Ke
  Wang}.} \bibinfo{year}{2018}\natexlab{}.
\newblock \showarticletitle{Personalized top-n sequential recommendation via
  convolutional sequence embedding}. In \bibinfo{booktitle}{\emph{Proceedings
  of the Eleventh ACM International Conference on Web Search and Data Mining}}.
  \bibinfo{pages}{565--573}.
\newblock


\bibitem[\protect\citeauthoryear{Vaswani, Shazeer, Parmar, Uszkoreit, Jones,
  Gomez, Kaiser, and Polosukhin}{Vaswani et~al\mbox{.}}{[n.\,d.]}]%
        {vaswani2017attention}
\bibfield{author}{\bibinfo{person}{Ashish Vaswani}, \bibinfo{person}{Noam
  Shazeer}, \bibinfo{person}{Niki Parmar}, \bibinfo{person}{Jakob Uszkoreit},
  \bibinfo{person}{Llion Jones}, \bibinfo{person}{Aidan~N Gomez},
  \bibinfo{person}{Lukasz Kaiser}, {and} \bibinfo{person}{Illia Polosukhin}.}
  \bibinfo{year}{[n.\,d.]}\natexlab{}.
\newblock \showarticletitle{Attention is all you need}.
\newblock  (\bibinfo{year}{[n.\,d.]}).
\newblock


\bibitem[\protect\citeauthoryear{Wang, Cao, Wang, Sheng, Orgun, and Lian}{Wang
  et~al\mbox{.}}{2021}]%
        {wang2021survey}
\bibfield{author}{\bibinfo{person}{Shoujin Wang}, \bibinfo{person}{Longbing
  Cao}, \bibinfo{person}{Yan Wang}, \bibinfo{person}{Quan~Z Sheng},
  \bibinfo{person}{Mehmet~A Orgun}, {and} \bibinfo{person}{Defu Lian}.}
  \bibinfo{year}{2021}\natexlab{}.
\newblock \showarticletitle{A survey on session-based recommender systems}.
\newblock \bibinfo{journal}{\emph{ACM Computing Surveys (CSUR)}}
  \bibinfo{volume}{54}, \bibinfo{number}{7} (\bibinfo{year}{2021}),
  \bibinfo{pages}{1--38}.
\newblock


\bibitem[\protect\citeauthoryear{Wang, Hu, Wang, Sheng, Orgun, and Cao}{Wang
  et~al\mbox{.}}{2019}]%
        {wang2019modeling}
\bibfield{author}{\bibinfo{person}{Shoujin Wang}, \bibinfo{person}{Liang Hu},
  \bibinfo{person}{Yan Wang}, \bibinfo{person}{Quan~Z Sheng},
  \bibinfo{person}{Mehmet Orgun}, {and} \bibinfo{person}{Longbing Cao}.}
  \bibinfo{year}{2019}\natexlab{}.
\newblock \showarticletitle{Modeling multi-purpose sessions for next-item
  recommendations via mixture-channel purpose routing networks}. In
  \bibinfo{booktitle}{\emph{International Joint Conference on Artificial
  Intelligence}}. International Joint Conferences on Artificial Intelligence.
\newblock


\bibitem[\protect\citeauthoryear{Wu, Yin, Lian, Yin, Gong, Zhou, and Yang}{Wu
  et~al\mbox{.}}{2021}]%
        {wu2021rethinking}
\bibfield{author}{\bibinfo{person}{Yongji Wu}, \bibinfo{person}{Lu Yin},
  \bibinfo{person}{Defu Lian}, \bibinfo{person}{Mingyang Yin},
  \bibinfo{person}{Neil~Zhenqiang Gong}, \bibinfo{person}{Jingren Zhou}, {and}
  \bibinfo{person}{Hongxia Yang}.} \bibinfo{year}{2021}\natexlab{}.
\newblock \showarticletitle{Rethinking Lifelong Sequential Recommendation with
  Incremental Multi-Interest Attention}.
\newblock \bibinfo{journal}{\emph{arXiv preprint arXiv:2105.14060}}
  (\bibinfo{year}{2021}).
\newblock


\bibitem[\protect\citeauthoryear{Xiao, Yang, Jiang, Wei, Hu, and Wang}{Xiao
  et~al\mbox{.}}{2020}]%
        {xiao2020deep}
\bibfield{author}{\bibinfo{person}{Zhibo Xiao}, \bibinfo{person}{Luwei Yang},
  \bibinfo{person}{Wen Jiang}, \bibinfo{person}{Yi Wei}, \bibinfo{person}{Yi
  Hu}, {and} \bibinfo{person}{Hao Wang}.} \bibinfo{year}{2020}\natexlab{}.
\newblock \showarticletitle{Deep Multi-Interest Network for Click-through Rate
  Prediction}. In \bibinfo{booktitle}{\emph{Proceedings of the 29th ACM
  International Conference on Information \& Knowledge Management}}.
  \bibinfo{pages}{2265--2268}.
\newblock


\bibitem[\protect\citeauthoryear{Xue, Dai, Zhang, Huang, and Chen}{Xue
  et~al\mbox{.}}{2017}]%
        {xue2017deep}
\bibfield{author}{\bibinfo{person}{Hong-Jian Xue}, \bibinfo{person}{Xinyu Dai},
  \bibinfo{person}{Jianbing Zhang}, \bibinfo{person}{Shujian Huang}, {and}
  \bibinfo{person}{Jiajun Chen}.} \bibinfo{year}{2017}\natexlab{}.
\newblock \showarticletitle{Deep Matrix Factorization Models for Recommender
  Systems.}. In \bibinfo{booktitle}{\emph{IJCAI}}, Vol.~\bibinfo{volume}{17}.
  Melbourne, Australia, \bibinfo{pages}{3203--3209}.
\newblock


\bibitem[\protect\citeauthoryear{Yang, Zhang, Yu, and Wang}{Yang
  et~al\mbox{.}}{2013}]%
        {yang2013sentiment}
\bibfield{author}{\bibinfo{person}{Dingqi Yang}, \bibinfo{person}{Daqing
  Zhang}, \bibinfo{person}{Zhiyong Yu}, {and} \bibinfo{person}{Zhu Wang}.}
  \bibinfo{year}{2013}\natexlab{}.
\newblock \showarticletitle{A sentiment-enhanced personalized location
  recommendation system}. In \bibinfo{booktitle}{\emph{Proceedings of the 24th
  ACM conference on hypertext and social media}}. \bibinfo{pages}{119--128}.
\newblock


\bibitem[\protect\citeauthoryear{Ye, Yin, and Lee}{Ye et~al\mbox{.}}{2010}]%
        {ye2010location}
\bibfield{author}{\bibinfo{person}{Mao Ye}, \bibinfo{person}{Peifeng Yin},
  {and} \bibinfo{person}{Wang-Chien Lee}.} \bibinfo{year}{2010}\natexlab{}.
\newblock \showarticletitle{Location recommendation for location-based social
  networks}. In \bibinfo{booktitle}{\emph{Proceedings of the 18th SIGSPATIAL
  international conference on advances in geographic information systems}}.
  \bibinfo{pages}{458--461}.
\newblock


\bibitem[\protect\citeauthoryear{Ye, Wang, Chen, Wang, Qin, and Yin}{Ye
  et~al\mbox{.}}{2020}]%
        {ye2020time}
\bibfield{author}{\bibinfo{person}{Wenwen Ye}, \bibinfo{person}{Shuaiqiang
  Wang}, \bibinfo{person}{Xu Chen}, \bibinfo{person}{Xuepeng Wang},
  \bibinfo{person}{Zheng Qin}, {and} \bibinfo{person}{Dawei Yin}.}
  \bibinfo{year}{2020}\natexlab{}.
\newblock \showarticletitle{Time matters: Sequential recommendation with
  complex temporal information}. In \bibinfo{booktitle}{\emph{Proceedings of
  the 43rd International ACM SIGIR Conference on Research and Development in
  Information Retrieval}}. \bibinfo{pages}{1459--1468}.
\newblock


\bibitem[\protect\citeauthoryear{Yuan, Karatzoglou, Arapakis, Jose, and
  He}{Yuan et~al\mbox{.}}{2019}]%
        {yuan2019simple}
\bibfield{author}{\bibinfo{person}{Fajie Yuan}, \bibinfo{person}{Alexandros
  Karatzoglou}, \bibinfo{person}{Ioannis Arapakis}, \bibinfo{person}{Joemon~M
  Jose}, {and} \bibinfo{person}{Xiangnan He}.} \bibinfo{year}{2019}\natexlab{}.
\newblock \showarticletitle{A simple convolutional generative network for next
  item recommendation}. In \bibinfo{booktitle}{\emph{Proceedings of the Twelfth
  ACM International Conference on Web Search and Data Mining}}.
  \bibinfo{pages}{582--590}.
\newblock


\bibitem[\protect\citeauthoryear{Zhou, Li, and Liang}{Zhou
  et~al\mbox{.}}{2020}]%
        {zhou2020cnn}
\bibfield{author}{\bibinfo{person}{Xiaokang Zhou}, \bibinfo{person}{Yue Li},
  {and} \bibinfo{person}{Wei Liang}.} \bibinfo{year}{2020}\natexlab{}.
\newblock \showarticletitle{CNN-RNN based intelligent recommendation for online
  medical pre-diagnosis support}.
\newblock \bibinfo{journal}{\emph{IEEE/ACM Transactions on Computational
  Biology and Bioinformatics}} (\bibinfo{year}{2020}).
\newblock


\bibitem[\protect\citeauthoryear{Zhu, Li, Liao, Wang, Guan, Liu, and Cai}{Zhu
  et~al\mbox{.}}{[n.\,d.]}]%
        {zhu2017next}
\bibfield{author}{\bibinfo{person}{Yu Zhu}, \bibinfo{person}{Hao Li},
  \bibinfo{person}{Yikang Liao}, \bibinfo{person}{Beidou Wang},
  \bibinfo{person}{Ziyu Guan}, \bibinfo{person}{Haifeng Liu}, {and}
  \bibinfo{person}{Deng Cai}.} \bibinfo{year}{[n.\,d.]}\natexlab{}.
\newblock \showarticletitle{What to Do Next: Modeling User Behaviors by
  Time-LSTM}.
\newblock  (\bibinfo{year}{[n.\,d.]}).
\newblock


\end{thebibliography}

\end{document}